\documentclass[aps,pra,superscriptaddress,reprint]{revtex4-1}
\usepackage[colorlinks=true,citecolor=blue,linkcolor=magenta]{hyperref}
\usepackage{graphicx}
\usepackage{bm}
\usepackage{dsfont}
\usepackage{amsmath,amssymb}
\usepackage{color,psfrag}
\usepackage{pstool}
\usepackage{array}
\usepackage{longtable}
\usepackage{wrapfig}
\usepackage{bbm}
\usepackage{color}

\newcommand{\bra}[1]{\langle #1|}
\newcommand{\ket}[1]{|#1\rangle}
\newcommand{\ex}[1]{\langle #1 \rangle}
\newcommand{\xo}[0]{\hat{z}}

\newcommand{\e}[0]{{\rm e}}
\newcommand{\tr}[0]{{\text{tr}}}

\usepackage{multirow}

\begin{document}
\title{Testing quantum gravity by nanodiamond interferometry with nitrogen-vacancy centers}
\author{Andreas Albrecht}
\affiliation{Institut f\"ur Theoretische Physik, Albert-Einstein-Allee 11, Universit\"at Ulm, 89069 Ulm, Germany}
\affiliation{Center for Integrated Quantum Science and Technology, Universit\"at Ulm, 89069 Ulm, Germany}
\author{Alex Retzker}
\affiliation{Racah Institute of Physics, The Hebrew University of Jerusalem, Jerusalem 91904, Israel}
\author{Martin B Plenio}
\affiliation{Institut f\"ur Theoretische Physik, Albert-Einstein-Allee 11, Universit\"at Ulm, 89069 Ulm, Germany}
\affiliation{Center for Integrated Quantum Science and Technology, Universit\"at Ulm, 89069 Ulm, Germany}

\begin{abstract}
Interferometry with massive particles may have the potential to explore the limitations of 
standard quantum mechanics in particular where it concerns its boundary with general
relativity and the yet to be developed theory of quantum gravity. This development is hindered 
considerably by the lack of experimental evidence and testable predictions. Analyzing effects 
that appear to be common to many of such theories, such as a modification of the energy 
dispersion and of the canonical commutation relation within the standard framework 
of quantum mechanics, has been proposed as a possible way forward. 
Here we analyze in some detail the impact of a modified energy-momentum dispersion in a Ramsey-Bord\'e setup and provide achievable bounds of these correcting terms when operating such an interferometer with nanodiamonds. Thus, taking thermal and gravitational disturbances into account will show that without specific prerequisites, quantum gravity modifications may in general be suppressed requiring a revision of previously estimated bounds. As a possible solution we propose a stable setup which is rather insensitive to these effects. Finally, we address the problems of decoherence and pulse errors in such setups and discuss the scalings and advantages with increasing particle mass.
\end{abstract}
\maketitle

A framework, unifying classical general relativity with quantum mechanics,  remains a crucial scientific challenge to date. As incompatibilities hinder the straightforward inclusion of gravity into the standard framework of quantum field theory, the development of a new theory, the quantum gravity, seems essential.  Spacetime quantization is a natural ingredient of such a theory stemming from its dynamical nature in general relativity. This has led to the notion of minimal length- and maximal energy scales\,\cite{hagar13, hossenfelder13}, commonly ascribed to the Planck-scales:  The Planck-length $l_p$ as the length where the Compton radius of quantum mechanics meets the Schwarzschild equivalent of gravitation $l_p=\sqrt{G\hbar/(c^3)}\simeq 1.6\cdot 10^{-35}\,m$ and the Planck mass $M_p=\hbar/(c\,L_p)=\sqrt{\hbar\,c/G}\simeq 2.1\cdot 10^{-8}\,{\rm kg}$. More abstract, these scales arise in a  combination of three fundamental constants, thereby forming new quantities that may or may not be of fundamental importance in  nature. This minimal lengthscale plays a crucial role in candidates for quantum gravity theories\,\cite{hagar13} such as string theory\,\cite{giddings13, blau09}, loop quantum gravity\,\cite{ashtekar04}, doubly special relativity\,\cite{magueijo02, amelino02} and in the field of black hole physics\,\cite{maggiore93}. The complete frameworks however are rather complex and incomplete in their physical interpretation. It has therefore been proposed to test common impacts of a spacetime quantization on standard quantum mechanics instead, such as the  modification of the energy dispersion relation\,\cite{amelino09} or the change of quantum mechanical commutation relations\,\cite{das08, ali11, pikovski12, hossenfelder13}. Incorporating such `universal' effects into existing frameworks inspired the proposal of numerous verification experiments both in the relativistic and non-relativistic regime. This has led to bounds on the magnitude of the anticipated fundamental scales, though, due to the smallness of the effects, an existence verification is still pending. In particular tests have been proposed within the framework of quantum optics as constraining the energy dispersion relation in atom interferometers\,\cite{amelino09, mercati10}, commutator measurements on nanomechanical oscillator systems\,\cite{pikovski12}, observation of energy level shifts and modified tunnelling rates\,\cite{ali11, das08}, holographic noise measurements\,\cite{hogan12, berchera13} and the appearance of modified photon scattering rates\,\cite{bekenstein12}. 

Here we will focus on the approach developed in\,\cite{amelino09, mercati10}, that proposes the test of quantum gravity induced energy dispersion modifications in the context of atom interferometry\,\cite{chu01, kasevich91}. This interferometry method, which relies on mapping acquired phase shifts to a well-controlled two-level quantum system,  allows for the ultraprecise detection of spatial potentials and accelerations. Accordingly its applications are versatile, ranging from gravitational constant\,\cite{kasevich91, kasevich92} and recoil based fine structure measurements\,\cite{wicht02, bouchendira11} to the detection of magnetic field gradients and gravitational waves\,\cite{dimopoulos08}. Replacing atoms by more massive particles is both interesting for testing the fundamental limits of quantum mechanics\,\cite{pikovski12, amelino09} and in the analysis of decoherence effects\,\cite{hornberger12, hackermuller04}. Of particular importance for the upcoming analysis, larger masses might be capable of enhancing Planck scale corrections. Color centers in diamond\,\citep{aharonovich11}, such as nitrogen- or silicon-vacancy centers, form promising candidates for that task.  As a specific example, we will focus on nanodiamonds comprising a nitrogen-vacancy center (NV$^-$-center) (see figure\,\ref{b_setup1}), that does provide a well-controlled internal level-structure along with the possibility for optical initialization and readout of the spin qubit ground states\,\cite{jelezko04}. Extraordinary long electronic coherence times up to milliseconds\,\cite{balasubramanian09} even at room temperature and combinations with dynamical decoupling methods\,\cite{delange10, *naydenov11, *cai12} make them promising candidates as a solid state qubit. Furthermore, optical trapping\,\cite{geiselmann13, neukirch13} and optical ground state manipulation\,\cite{yale13,hilser12} have been demonstrated, which form important prerequisites for recoil based interferometry experiments. The interferometric application of nanodiamonds in a trapped configuration has been proposed only recently\,\cite{yin13,scala13}. 

This paper is organized as follows: We will start by introducing the modified energy dispersion relation along with the interferometric setup for its measurement in section\,\ref{sect_su}. Based on the interference signal and assuming various sources of noise and imprecisions, we then provide precision bounds in section\,\ref{sect_pb} that may be anticipated in the determination of the quantum gravity modifications. Here we focus in particular on the mass scaling and a comparison between atomic and nanoparticle setups. Having identified the contribution for which the verification of their existence or non-existence seems feasible with nanodiamonds,  we analyze the interference signal under the influence of gravitation and thermal motion in section\,\ref{sect_oo}, showing that this contribution is in general unobservable as a result of suppression and decoherence mechanisms. As a viable solution we then provide in section\,\ref{sect_stab} a modified robust setup based on momentum inversion by gravitation, allowing to restore the perfect phase and visibility, that otherwise could just be achieved under challenging conditions. Whereas the preceding analysis is based on a particular closed path interference contribution of two selected paths, section\,\ref{sect_totint} is dedicated to extend that concept to all paths involved in the setup.   In section\,\ref{sect_obs} we address the question whether massive particles and quantum optical schemes are suitable for tests of quantum gravitational corrections, thereby pointing out the still unresolved controversies that have emerged in the literature. 
Practical consideration as the combination with decoupling sequences, spatial decoherence,  visibility reduction by imperfect pulses, time and pulse errors are discussed in section\,\ref{sect_pract}, focusing again on the particularities of nanoparticle setups. Following the main text, Appendix\,\ref{append_1} provides the derivation of the interferometer phase at the example of a specific path combination and based on a recently developed operator formalism\,\cite{schleich2013}. In Appendix\,\ref{append_overlap} the coherence matrix element as required for the interference term evaluation is calculated for a thermal harmonic oscillator state. Appendix\,\ref{append_2} gives an analysis of the developed robust stability setup and the required momentum conditions, and Appendix\,\ref{append_3} discusses the difference of the modified energy dispersion approach to studies based on a modified commutation relation.

\section{Modified energy dispersion relation and interferometer phase}\label{sect_su}
\begin{figure}[htb]
\begin{centering}
\includegraphics[scale=0.4]{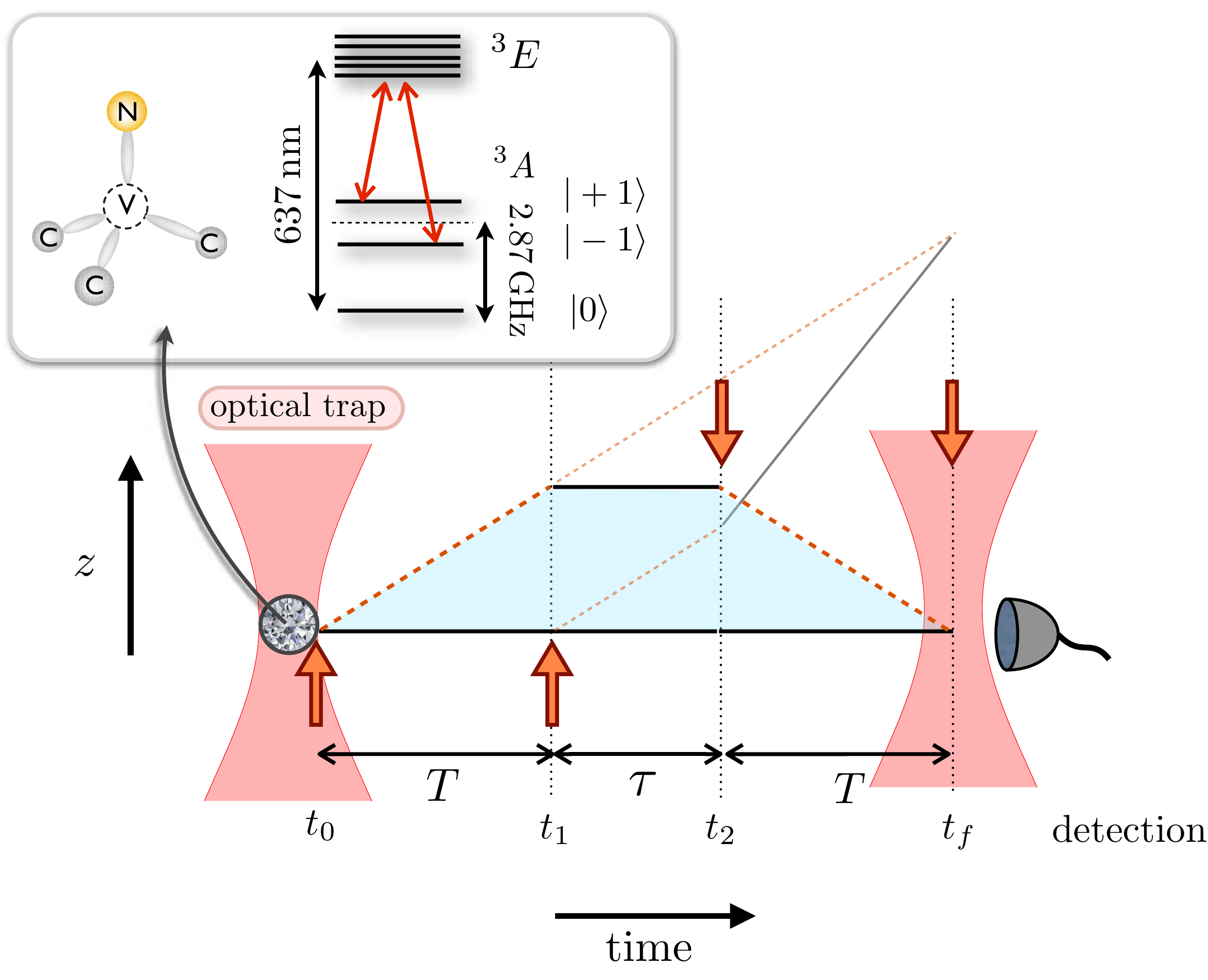}
\caption{\label{b_setup1} (Color online) \textbf{Ramsey-Bord\'e interferometer setup and NV-center energy levels}. An initially trapped nanoparticle undergoes the illustrated interferometer sequence where black lines correspond to the NV being in the ground state $\ket{g}$ (e.g. $\ket{-1}$ or $\ket{0}$) and red dashed lines to the excited state $\ket{e}$ (e.g. $\ket{+1}$). Red arrows indicate $\pi/2$-laser Raman transitions, beam splitter operations with recoil transfer $\hbar k$ in the corresponding laser directions. Those transitions can be realized using a $\Lambda$-scheme between the $\ket{+1}$, $\ket{-1}$\,\cite{togan11} or $\ket{0}, \ket{+1}$\,\cite{yale13} $^3A$ ground states and an appropriate excited state of the manifold\,$^3E$ with a photon recoil momentum $\hbar k\simeq 2\hbar\cdot 2\pi/(637\,{\rm nm})$. The inset depicts the nitrogen vacancy center energy level structure. }
\end{centering}
\end{figure}
The approach developed in\,\cite{amelino09, mercati10} proposes the analysis of a modified energy dispersion, that in the non-relativistic limit takes the the form
\begin{equation}\label{pqg1} E(p)=\frac{p^2}{2\,m}+\xi_1\,\frac{m\,c\,p}{2\,M_p}+\xi_2\,\frac{p^2}{2\,M_p}  \end{equation}
with $p$ the particle momentum, $m$ the mass and $c$ the speed of light. $\xi_1$ and $\xi_2$ form free parameters, that have to be constrained by experiments. If the Planck scale plays a fundamental role both parameters are expected to be of order one. Indications for such linear and quadratic corrections in the particle momentum can be found in loop quantum gravity\,\cite{alfaro00} and doubly special relativity approaches\,\cite{amelino02}, respectively. More abstract, the modifications induced can be considered as the first order correction in the Planck mass of the form $(1/M_p)\,\Delta^{(1)}(p,m)$ under the assumption that $m$ still takes the role of the rest mass ($\Delta^{(1)}(p\to 0,m)=0$) and for $M_p\to\infty$ the standard dispersion relation is recovered\,\cite{mercati10}. In the non-relativistic limit $p\ll mc$ the $\xi_1$ and $\xi_2$ contributions then form the leading and next to leading possible contributions to $\Delta^{(1)}(p,m)$, respectively. In a broader context, the general ansatz\,(\ref{pqg1}), and in particular the $\xi_2$ contribution, also serves as a test for Lorentz symmetry breaking\,\cite{mattingly05, amelino09}. 

As pointed out in\,\cite{amelino09, mercati10}, which is exclusively based on an analysis of energy-momentum conservation, the phase of a Ramsey-Bord\'e interferometer as illustrated in figure\,\ref{b_setup1}, corresponds  to the difference in the kinetic energy between the two paths. We will show in Appendix A, that for the closed paths combination illustrated in figure\,\ref{b_setup1} and a potential at most linear in position (leading to an inertial force), $H(\hat{p},\hat{x})=E(\hat{p})+V(\hat{x})$ with $V'(\hat{x})={\rm const.}$, the interferometer phase for an initial definite momentum state $\ket{\vec{p}}$ is given by
\begin{equation}\label{pqg2}\begin{split}  \phi_{\rm int}=&\frac{1}{\hbar}\,\left[\int_{t_0}^{t_1}\Delta E(\vec{p}-\nabla V(x)\,t',\hbar \vec{k})\mathrm{d}t'\right.\\&+\left.\int_{t_2}^{t_f}\Delta E(\vec{p}-\nabla V(x)\,t',-\hbar \vec{k})\mathrm{d}t'  \right]+\Delta\varphi \end{split}\end{equation} 
with the times defined as in figure\,\ref{b_setup1}, $\Delta E(\vec{p},\delta\vec{p})=E(|\vec{p}+\delta\vec{p}\,'|)-E(|\vec{p}|)$ and $\Delta\varphi$ the phase factor originating from the laser pulse interactions with $\Delta\varphi=\left[ \varphi(t_f)-\varphi(t_2)+\varphi(t_1)-\varphi(t_0) \right]+\int_{t_0}^{t_1}\delta(t')\,\mathrm{d}t'+\int_{t_2}^{t_f}\delta(t')\,\mathrm{d}t'$. Herein $\delta$ denotes the laser detuning and $\varphi(t)$ the absolute laser phase at time $t$. In the absence of an accelerating force, for equal time intervals $T$ and a constant laser phase and detuning $\delta$, this reduces to $\phi_{\rm int}=1/\hbar\,(\Delta E(\vec{p},\hbar \vec{k})+\Delta E(\vec{p},-\hbar\vec{k}))\,T+\delta\,T$. Upon completing the interferometer sequence, this phase is measurable as population oscillations contributing to the probability for finding the particle in the ground state as $p_g'=1/8(1+\cos(\phi_{\rm int}))$. The total probability, involving the influence of additional paths of the interferometric setup will be analyzed in section\,\ref{sect_totint}.

\section{Precision bounds on quantum gravity parameters}\label{sect_pb}

A variety of error sources and uncertainties, both of practical and fundamental nature, limit the achievable precision in the interferometric determination of the quantum gravity parameters $\xi_1$ and $\xi_2$. It will turn out that for a bound on $\xi_2$ no gain in precision can be expected in going for more massive particles such as nanodiamonds; from a practical perspective even the contrary holds true.  On the other hand, bounds on $\xi_1$ benefit considerably from increasing masses and already a rather small nanodiamond size could make its existence amenable to verification provided $\xi_1\sim 1$. However, both for atoms and nanoparticles, such a $\xi_1$ term will generally be suppressed by gravitational and thermal motion, the latter effect even scaling disadvantageously with increasing particle mass. Thus it is in general impossible to extract this data from existing interference experiments as has been proposed and performed in\,\cite{amelino09,mercati10}, therefore creating the need for a new setup and analysis as will be discussed in the upcoming sections.

The (perfect) external interferometric phase following out of\,(\ref{pqg2}) takes the form
\begin{equation}\label{bound1}\phi=\frac{\hbar k^2}{m}\,T+\xi_1\,\frac{m}{M_p}\,c\,k\,T+\xi_2\,\frac{\hbar k^2}{M_p}\,T\end{equation}
and we will refer to the first term, the phase of an unmodified energy dispersion, as the `zero order' contribution in what follows.  

\begin{figure}[htb]
\begin{centering}
\includegraphics[scale=0.5]{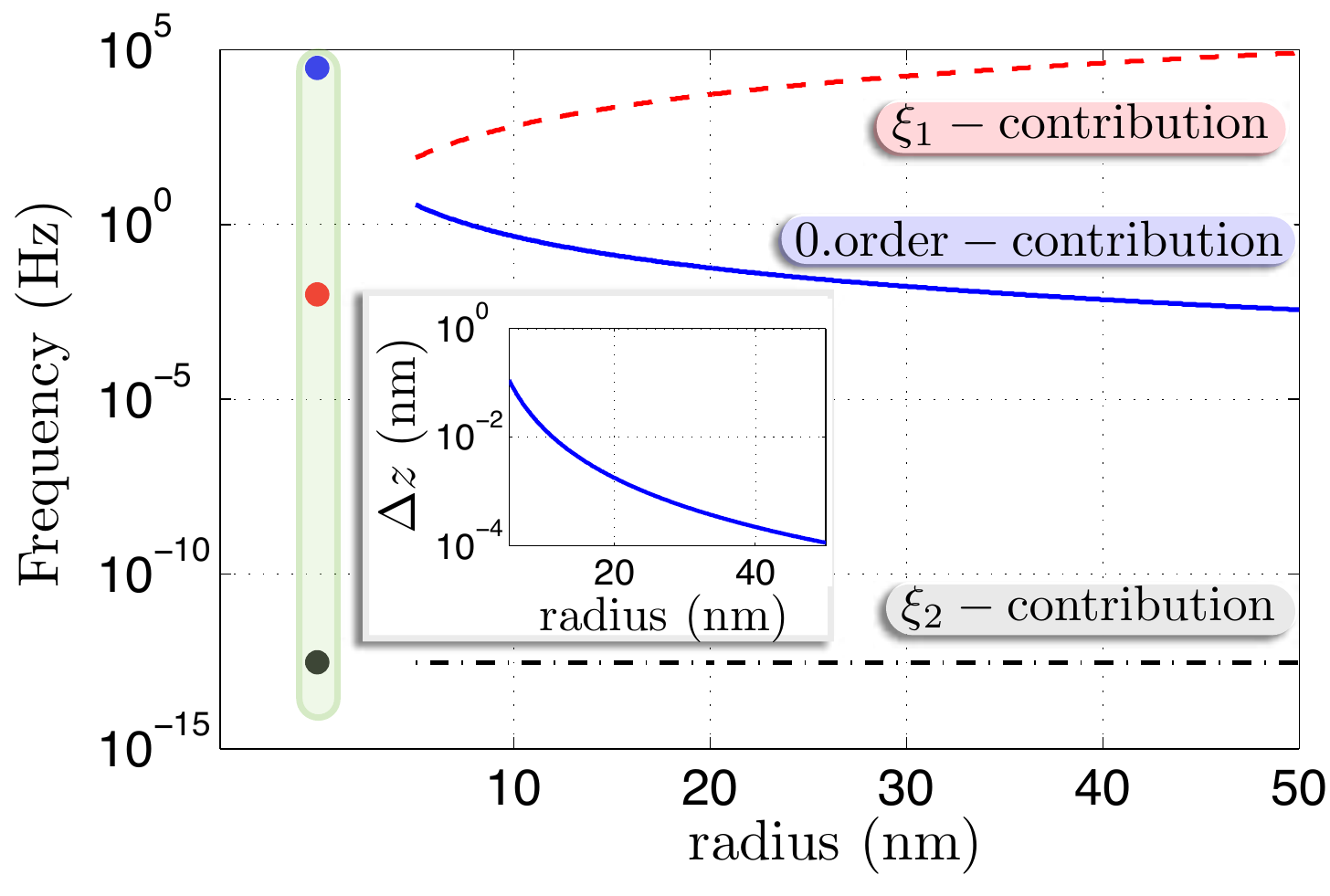}
\caption{\label{b_contribscale} (Color online) \textbf{Frequency contribution and momentum splitting.} Interferometry phase frequency for different nanodiamond radii: The unperturbed recoil phase (zero order contribution, \textit{blue solid}), the $\xi_1$ quantum gravity correction (\textit{red dashed}) and the $\xi_2$ correction (\textit{black dashed dotted}). The circles in the green shaded area (top to bottom: zero order, $\xi_1$, $\xi_2$) indicate the corresponding values for a Cs atom interferometer in the same colours. \textit{Inset: } Maximal spatial separation of the interferometer paths for an interferometer time $T=100\,\mu s$ and nanodiamonds.  }
\end{centering}
\end{figure}

\begin{figure}[htb]
\begin{centering}
\includegraphics[scale=0.36]{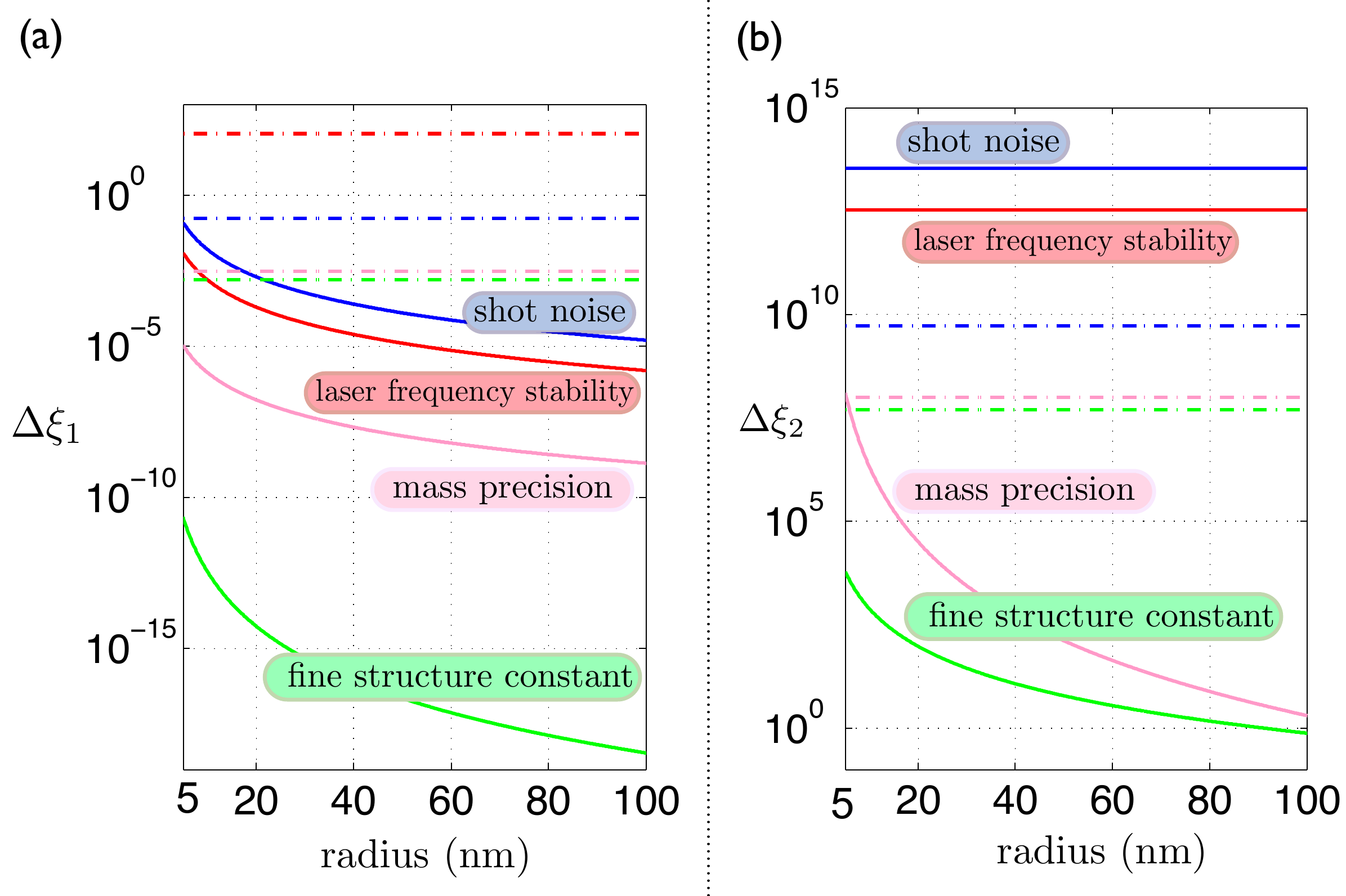}
\caption{\label{b_bounds} (Color online) \textbf{Precision bounds} on \textbf{(a)} $\xi_1$  and \textbf{(b)} $\xi_2$ arising from various error sources for different nanodiamond radii. Dashed dotted lines indicate the bounds for an atomic interferometry setup (Cs atoms). For the shot noise limitation we assume a number of $N=1000$ repetitions per data point and an interferometer time of $T=500\,\mu s$ for nanodiamonds, whereas for Cs atoms $N=10^8$  and $T=10\,{\rm ms}$ in accordance with recent atomic interference experiments\,\cite{bouchendira11} is assumed. The wavevector corresponds to a Raman transition recoil contribution with $k\sim 2\cdot 10^7 \,{\rm m^{-1}}$. The mass precision for Cs atoms is given by\,\cite{iupac13} $\Delta m_{\rm Cs}/m_{\rm Cs}=10^{-9}$, whereas for nanodiamonds a variance of $\Delta m_{\rm NV}=\pm m_{\rm C12}$ corresponding to one carbon atom is assumed. The $\Delta \left(\hbar/m_{e}\right)$ (fine structure) uncertainty has been extracted from electron magnetic moment anomaly measurements\,\cite{hanneke08, aoyama12} leading to $\Delta \left(\hbar/m_e\right)=5.9\cdot 10^{-14}\,{\rm Js/kg}$.  }
\end{centering}
\end{figure}    

The magnitudes of the different phase contributions are depicted in figure\,\ref{b_contribscale} for different particle masses. Note in particular that for nanodiamonds the $\xi_1$-term, provided that it exists, forms the dominant phase contribution, whereas this role is ascribed to the zero order phase contributions $\phi_0=(\hbar k^2/m)T$ for an atomic system. This turns out to be crucial for the error scaling in that uncertainties in the zero order phase term, denoted as `relative errors', play a significant role for atomic systems whereas they are much less prominent for nanodiamonds. In particular for constraining $\xi_1$ with nanoparticles, `relative errors' play a minor role compared to `absolute errors', the latter referring to mass and wavevector independent uncertainties such as shot noise or frequency imprecisions. The mass and wavevector scaling for these two different types is given in table\,\ref{tab_bounds} along with the scaling of different error sources, the latter ones plotted for realistic experimental parameters in figure\,\ref{b_bounds}. As a consequence of the different error scaling, the absolute error scaling $\Delta\xi_1\propto 1/(mk)$ and $\Delta\xi_2\propto 1/k^2$ favours larger recoil contributions for nanodiamonds, whereas the opposite conclusion can be drawn for atomic systems\,\cite{mercati10}  based on the relative error scaling $\Delta\xi_1\propto k/m^2$.

\begin{table*}[htb]
\begin{tabular}{|l l|c| c|}\hline
  \multicolumn{2}{|l|}{Shot noise limit} & $\Delta\xi_1^{\rm sn}=\frac{1}{\mathbf{\sqrt{N}}}\,\frac{M_p}{cT}\frac{1}{\mathbf{m\,k}}$ & $\Delta\xi_2^{\rm sn}=\frac{1}{\mathbf{\sqrt{N}}}\,\frac{M_p}{T\,\hbar\mathbf{k^2}}$ \\\hline
 \multicolumn{2}{|l|}{Laser frequency stability ($\Delta\delta$)}  & $\Delta\xi_1^{\rm lf}=\frac{M_p}{c}\,\frac{1}{\mathbf{m\,k}}\,\Delta\delta$ & $\Delta\xi_2^{\rm lf}=\frac{M_p}{\hbar}\frac{1}{\mathbf{k^2}}\,\Delta\delta$ \\\hline 
\multicolumn{2}{|l|}{Mass precision} & $\Delta\xi_1^{\rm mass}=\xi_1\frac{\Delta m}{\mathbf{m}}+\frac{M_p}{\mathbf{m^2}}\,\frac{\hbar \mathbf{k}}{c}\,\frac{\Delta m}{\mathbf{m}}$ & $\Delta\xi_2^{\rm mass}=\frac{M_p}{\mathbf{m}}\,\frac{\Delta m}{\mathbf{m}}$ \\\hline
\multicolumn{2}{|l|}{Precision of the fine structure constant / of ($\hbar/m_e$)} & $\Delta\xi_1^{\alpha}= \frac{M_p\,m_e}{c}\,\frac{\mathbf{k}}{\mathbf{m^2}}\,\Delta\left(\frac{\hbar}{m_e}\right)$ & $\Delta\xi_2^{\alpha}= \frac{M_p}{\mathbf{m}}\,\frac{1}{\hbar/m_e}\,\Delta\left(\frac{\hbar}{m_e}\right)$ \\\hline\hline
\multirow{2}{*}{General scaling} &\quad absolute value &  $\Delta\xi_1\propto 1/(m\,k)$  &  $\Delta\xi_2\propto 1/k^2$ \\
& \quad relative value (zero order term)  & $\Delta\xi_1\propto k/m^2$ & $\Delta\xi_2\propto 1/m$\\\hline
\end{tabular}
\caption{\label{tab_bounds}Imprecision bound formulas for the estimation of the parameters $\xi_1$ and $\xi_2$ for different sources of imperfection. The last row indicates the general scaling with the particle mass $m$ and the wavevector $k$ for errors that are due to fluctuations of the zero-order phase (relative value) and mass and wavevector independent errors (absolute value), respectively. }
\end{table*}

The $\xi_1$-bound benefits for both types of errors from an increasing mass. Strikingly, due to $\xi_1$ being the dominant phase contribution for nanodiamonds, already a rather small radius of 5\,nm ($\sim 10^6\,{\rm amu},\,\sim 9\cdot 10^{-14}M_p$) might allow for the existence or non-existence verification of such a $\xi_1$-term. The shot noise is, along with imprecisions in the laser detunings, the main and limiting error source, whereas fundamental knowledge of the fine structure constant will be very unlikely to limit the precision under realistic experimental conditions.

In contrast, the $\xi_2$-contribution exhibits a mass independent scaling for `absolute' errors. From a practical point of view this favours atoms over nanoparticles, the former benefiting from the ability to realise parallel interferometric setups of up to $\sim 10^8$ atoms\,\cite{bouchendira11}, long coherence times $T\sim 10\,{\rm ms}$ , higher mass precisions originating in their elementary particle nature and elaborated techniques for large momentum transfers\,\,\cite{chiow09, chiow11, bouchendira11}.  In particular the lack of parallelism for nanoparticle setups forms a challenging obstacle in overcoming a predominant shot noise error.  Only for the much smaller relative errors can a favorable scaling with particle mass be expected.   In any case, a tight $\xi_2$-estimation remains a challenging task as any frequency imprecisions, e.g. the laser detuning stability, have to be compared to the very small $\propto 10^{-13}\,{\rm Hz}$ $\xi_2$-contribution frequency  (see figure\,\ref{b_contribscale}). In previous work based on the experimental data of a Cs atom interferometer\,\cite{wicht02}, the $\xi_2$ parameter has been constrained to\,\cite{mercati10} $\Delta\xi_2\simeq 2.6\cdot 10^9$, and an improvement of one order of magnitude $\Delta\xi_2\simeq 3\cdot 10^8$ can be obtained from more recent experimental data based on Rb-atoms\,\cite{bouchendira11}, still far from verifying or falsifying the existence of such a quantum gravity correction.

\section{Can a $\xi_1$-term be observed in experiments?}\label{sect_oo}
As outlined in the previous section, it is merely the $\xi_1$ contribution that may gain advantage from an increased mass, in addition, it is the quantity for which the verification of its  existence or non-existence seems feasible. However this term exhibits two crucial drawbacks:  First, there remains in general a dependence on the particle momentum $\vec{p}$ even for a perfect interferometer sequence. This resembles the situation of open interferometers, in which by averaging over an initial thermal distribution of particle momenta, decoherence, and therefore a reduction of the interferometer visibility, may be anticipated.  Second, contrary to the phase contributions quadratic in momentum,  the impact of gravitation on the linear $\xi_1$ term does not appear as a separate phase factor nor can it be eliminated by choosing the recoil momentum orthogonal to the gravitation direction.  In contrast, pure gravitation will suppress the $\xi_1$ phase term contribution. Therefore, in the absence of a specific preparation of the interferometer sequence, the coherent $\xi_1$ contribution appearing in\,(\ref{bound1}) will in general be unobservable. Whereas the thermal phase suppression scales $\propto\sqrt{m}$, thus leading to more restrictive prerequisites for increasing mass, the equally challenging gravitational counterpart turns out to be mass independent. In the following we will begin by analyzing the phase term for a constant non-zero initial momentum, which will lead to a description of the phase-term behaviour for a thermal particle along with conditions for its observation. Next, the influence of gravitation will be considered in more detail with a subsequent discussion of the influence of both gravitation and thermal motion.

\subsubsection{$\xi_1$-phase and suppression for a constant momentum}
The external phase contribution of the $\xi_1$-term according to (\ref{pqg1}) and (\ref{pqg2}) for the lower closed path interference contribution, follows as
\begin{equation}\label{xi2}\begin{split} \phi_{\xi_1}=\mu &\left(\,\int_{t_0}^{t_1} \left| \vec{p}_1(t')+\hbar\,\vec{k}\right|-\left|\vec{p}_1(t')  \right|\,\mathrm{d}t'\right.\\
&\left.\,+\int_{t_2}^{t_f} \left| \vec{p}_2(t')-\hbar\,\vec{k}\right|-\left|\vec{p}_2(t')  \right|\,\mathrm{d}t'\right)
  \end{split} \end{equation} 
with $\mu=\xi_1m\,c/(2\hbar\,M_p)$, the times as defined in figure\,\ref{b_setup1} and we will assume that $t_1-t_0=t_f-t_2=T$.  As a special, and particularly relevant case, we will consider the  situation of a constant momentum in each of the two intervals, but allow for different momenta $\vec{p}_1$ and $\vec{p}_2$ in the first ($[t_0,t_1]$) and second ($[t_2,t_f]$) interferometer cycle, respectively.

\begin{figure}[htb]
\begin{centering}
\includegraphics[scale=0.5]{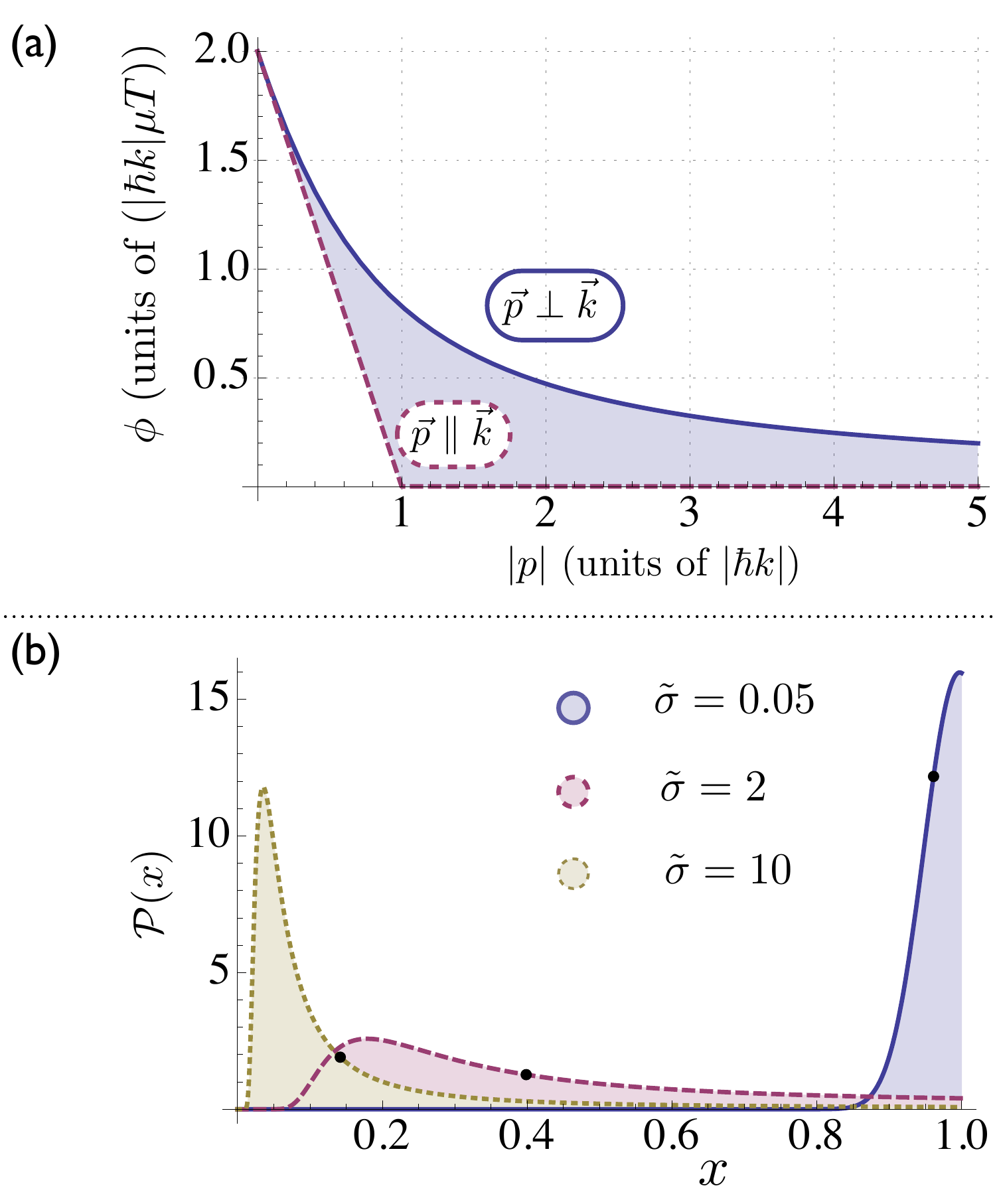}
\caption{\label{b_phasevsp} (Color online) \textbf{Momentum phase suppression and frequency distribution.} \textbf{(a)} Suppression of the $\xi_1$-phase contribution for an initial non-zero momentum $\vec{p}$ (constant during the interferometric cycle) and different orientations to the recoil contribution $\hbar \vec{k}$. Intermediate orientations do fall into the blue shaded region. \textbf{(b)} $\xi_1$-term frequency distribution of the dimensionless frequency $x=\omega/(2\mu\hbar k)$ for a Gaussian momentum distribution with different variances $\tilde{\sigma}=\sigma/(\hbar k)$ and assuming $\vec{p}\perp \hbar\vec{k}$.}
\end{centering}
\end{figure}

Let us first consider the case of equal momenta $\vec{p}_1=\vec{p}_2=\vec{p}$ as illustrated in figure\,\ref{b_phasevsp}\,(a). Such a situation describes a fixed particle momentum up to the beam splitter induced modifications, as might correspond to a particular momentum out of a thermal distribution.  With increasing magnitude of that initial momentum, the $\xi_1$ phase term reveals a purely decaying behaviour. Thus, a significant phase contribution in that regime can be expected exclusively in the limit $|p|\ll|\hbar k|$. Such a decay behaviour with the momentum generally characterizes the $\xi_1$-phase, which will turn out to make the phase observation challenging.  

Originating in the rather small recoil induced momentum shift $\hbar k$, that in most cases is much smaller than the thermal momentum variance or the momentum gain by gravitation, the limiting regime $|p_1|,|p_2|\gg\hbar\,k$ will be of particular importance. In that case, $|\vec{p}_i(t)+\hbar\vec{k}|\simeq |\vec{p}_i(t)|\,(1+\hbar k/|\vec{p}_i(t)|\,\cos\theta_i+|\hbar k|^2/(2\,|\vec{p}_i(t)|^2)\,[1-\cos^2\theta_i]) $, the phase (\ref{xi2}) can be approximated by
\begin{equation}\begin{split}\label{xi3}   \phi_{\xi_1}\simeq &\mu\,T\,\left\{ |\hbar k|\left(\cos\theta_1-\cos\theta_2 \right)\right.\\ &\left.+\frac{|\hbar k|^2}{2}\,\left[(1-\cos^2\theta_1)\frac{1}{|p_1|}+(1-\cos^2\theta_2)\,\frac{1}{|p_2|} \right]\right\} \end{split}\end{equation}   
where $\cos\theta_i=p_i^z/|p_i|\,\text{sgn}(\hbar{k})$ with $\theta_i$ the angle between the momentum $\vec{p}_i$ and the recoil z-direction. The second contribution describes the recoil components orthogonal to the momentum direction and is purely decaying with the momentum magnitude. This holds at each instant of time and leads to a significant suppression of that term. In contrast, the first term will vanish as a result of the phase cancellation in the first and second interferometric cycle unless there occurs a change in the momentum direction, that is, unless $\theta_1\neq\theta_2$. This term corresponds to the momentum parallel recoil contribution.   

Two important consequences follow out of the observed behaviour in the limit $p\gg\hbar k$: First,  in the absence of any directional momentum change ($\theta_1=\theta_2={\rm const.}$) or for a momentum purely orthogonal to the recoil direction, the $\xi_1$-phase contribution will decay with increasing momentum. We will see in the following sections that, except for specifically designed setups, this renders the phase observation essentially unobservable under the influence of gravitation or a thermal momentum distribution. Second, for the recoil being sufficiently parallel to the particle momentum and a momentum inversion in between the first and second interferometric cycle, a significant phase contribution can be expected. This latter observation will lead us to the creation of a `stability setup' in section\,\ref{sect_stab}, which will turn out to recover the unperturbed phase in the presence of gravitation and thermal motion.  \par

\subsubsection{Thermal momentum distribution}\label{sect_thermmom}
We will now assume the initial momentum of the interferometric particle being thermally distributed. Correspondingly the final population will have to be averaged over that distribution of initial momenta, as will be described in more detail in section\,\ref{sect_totint}. From a different point of view, this corresponds to an average over the frequencies $\omega(p)$ of the $\xi_1$ phase contribution defined by $\phi_{\xi_1}=\omega(p)\,T$. 

Two important consequences arise from such a thermal distribution: First, as described in the preceding section and as can be deduced from (\ref{xi3}),  the suppression of $\omega(p)$ with increasing $p$, i.e. the average frequency decreases with an increasing momentum variance.  Second, the momentum variance will be reflected in the frequency variance $\omega(p)$, the latter being responsible for a coherence decay, a decay of the corresponding interference term and a reduction of the coherence time $T_2$.

We will assume the initial momentum to be Gaussian distributed. This will be the case for an initial thermal harmonic oscillator state as discussed in Appendix\,\ref{append_overlap}, which leads to a zero-mean distribution with variance $\sigma^2=(\hbar/2)\,m\omega_{\alpha}(1+2\,\ex{\hat{n}})$.  Herein the thermal population $\ex{\hat{n}}=(\text{exp}(\hbar\omega_{\alpha}/(k_B\,\mathcal{T})-1))^{-1}$ and $\omega_\alpha$ the oscillator frequency in the corresponding spatial direction $\alpha\in x,y,z$. This reduces to the Boltzmann distribution $\sigma^2\simeq m\,k_B\,\mathcal{T}$ for $k_B\,\mathcal{T}\gg\hbar\omega_{\alpha}$ or equivalently for a free particle. In any limit the momentum variance, responsible for the phase suppression, grows with increasing mass $\sigma\propto \sqrt{m}$, and consequently makes a phase observation more challenging for massive particles. 

Figure\,\ref{b_phasevsp}\,(b) illustrates the corresponding interference phase frequency distribution $\mathcal{P}(\omega(p))$ out of the momentum distribution for different variances $\tilde{\sigma}=\sigma/(\hbar\,k)$. Albeit this assumes the momentum orthogonal to the recoil direction, a similar behaviour holds for a parallel configuration. This allows to identify three different regimes: \par For $|\sigma|\ll |\hbar k|$ the mean frequency approaches the optimal phase $\omega\simeq 2\mu\,\hbar k$ as given in\,(\ref{bound1}). As the frequency variance around this optimal frequency is small, decoherence only has a negligible influence. \par In contrast,  for $|\sigma|\simeq |\hbar k| $ all possible frequencies $\omega\in [0,2\mu\,\hbar k]$ appear with almost equal probability, therefore leading to the largest frequency variance and a decay of interference fringes. Consequently, coherent phase contributions cannot be observed in that regime. \par Last, in the limit $|\sigma|\gg|\hbar k|$ the phase suppression with increasing momentum is responsible for a dominant frequency contribution $\omega\simeq 0$.  As only very few momentum states have significant non-zero frequencies, the frequency variance decreases again. Thus no coherent $\xi_1$-oscillations can be observed in that regime, but at the same time decoherence decreases with increasing $\sigma$, therefore not disturbing the observation of other phase contributions on that timescale (e.g. the zero-order contribution).

\begin{figure}[bth]
\begin{centering}
\includegraphics[scale=0.32]{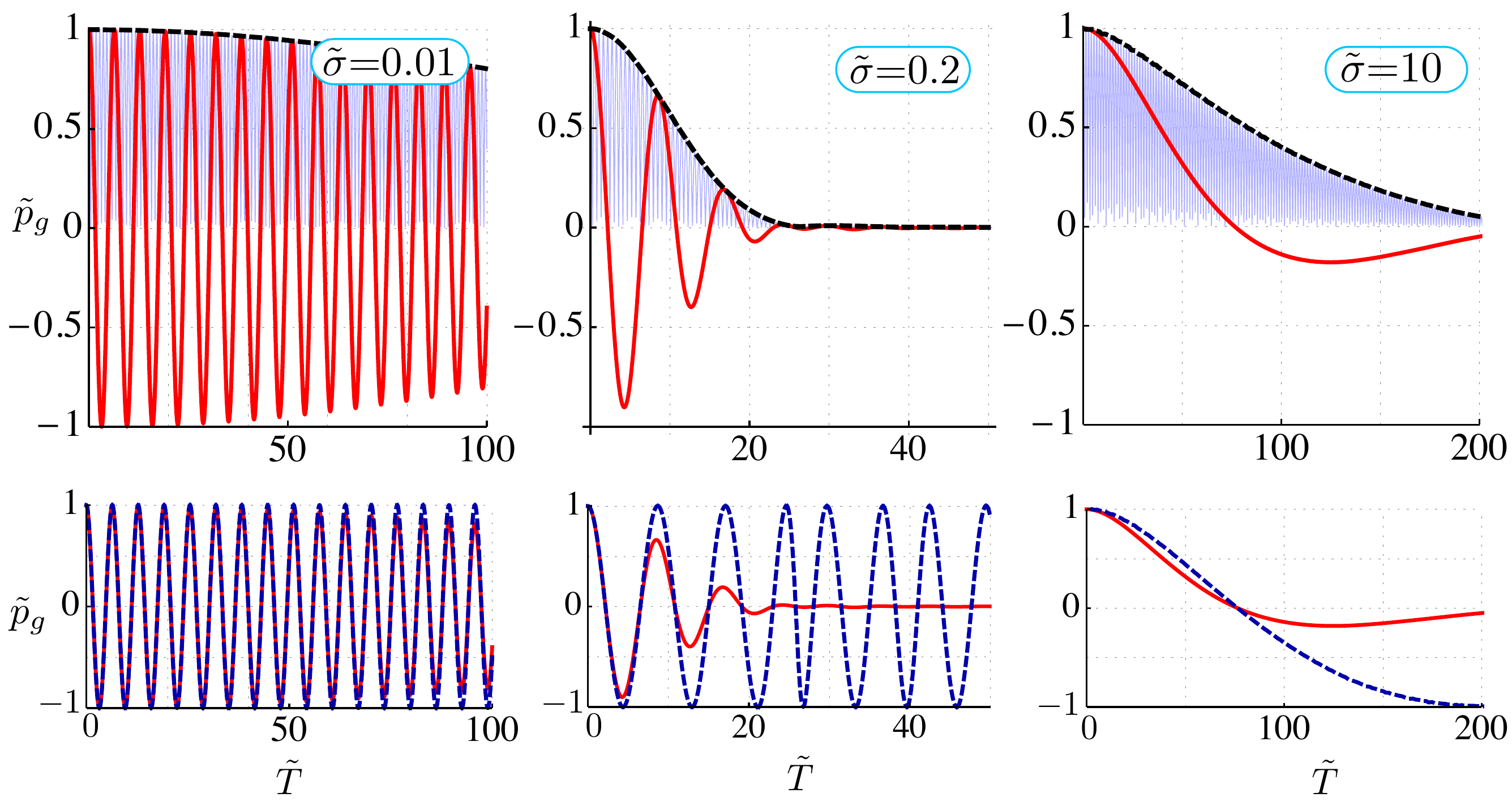}
\caption{\label{b_thermaldecay} (Color online) \textbf{Thermal influence on the $\xi_1$-contribution.} $\xi_1$-interference contribution \textit{(red solid)} $p_g'=(1/8)\,\left(1+\tilde{p}_g \right)$ for different values of the thermal momentum variance $\tilde{\sigma}=\sigma/(\hbar k)$ plotted versus the dimensionless time $\tilde{T}=2\mu (\hbar k) T$. A fast oscillation frequency, whose absolute value contribution is plotted here, has been added artificially to the phase \textit{(light blue)} to extract the decoherence decay envelope \textit{(black dashed)}. The lower panel shows the coherent phase contribution \textit{(blue dashed)} with subtracted decay envelope and the actually observed $\xi_1$-interference contribution \textit{(red solid)}.  }
\end{centering}
\end{figure}

\begin{table}[tbh]
\begin{tabular}{|l r|c| c|c|}\hline
& & r=5\,nm & r=50\,nm & Cs atom\\\hline\hline
HO ground state & $\tilde{\sigma}$  & 0.37  & 12  & $4\cdot 10^{-3}$ \\ 
($\omega=2\pi\cdot 1\,{\rm Hz}$)				& $T_2 / {\rm s}$   &0.03    &$2.5\cdot 10^{-4}$ &$9\cdot 10^3$ \\\hline
\multirow{2}{*}{$\mathcal{T}$=4\,$\mu$K } & $\tilde{\sigma}$  & 153  & $5\cdot 10^3$  & $1.7$ \\ 
				        & $T_2 / {\rm s}$   &3.2    & 0.1 &$3.6\cdot 10^2$ \\\hline
\multirow{2}{*}{$\mathcal{T}$=1\,mK } & $\tilde{\sigma}$  & $2\cdot 10^3$  & $8\cdot 10^4$  & $27$ \\ 
				      & $T_2 / {\rm s}$   &52    & 1.6 &$4.5\cdot 10^3$ \\\hline
\multirow{2}{*}{$\mathcal{T}$=10\,K } & $\tilde{\sigma}$  & $2\cdot 10^5$  & $8\cdot 10^6$  & $2.6\cdot 10^3$ \\ 
				      & $T_2 / {\rm s}$   &$5\cdot 10^3$   & $1.6\cdot 10^2$ &$4.7\cdot 10^5$ \\\hline
\end{tabular}
\caption{\label{t_sigval}\textbf{Thermal momentum variance and $T_2$-time.} Thermal momentum variance $\tilde{\sigma}=\sigma/(\hbar k)$ for $k=1.9\cdot 10^7\,m^{-1}$ corresponding to a $637\,{\rm nm}$ Raman transition and the corresponding $T_2$-time for different nanodiamond radii / a Cs atom and different temperatures. The first row corresponds to the quantum fluctuations of an harmonic oscillator (HO) ground state with frequency $\omega$. Note that the variance, which determines the regime as outlined in the main text, scales $\propto \sqrt{m}$, whereas the absolute timescale is proportional to $1/m$ reflecting the mass dependence of the $\xi_1$-contribution term. }
\end{table}

\begin{figure*}[htb]
\begin{centering}
\includegraphics[scale=0.45]{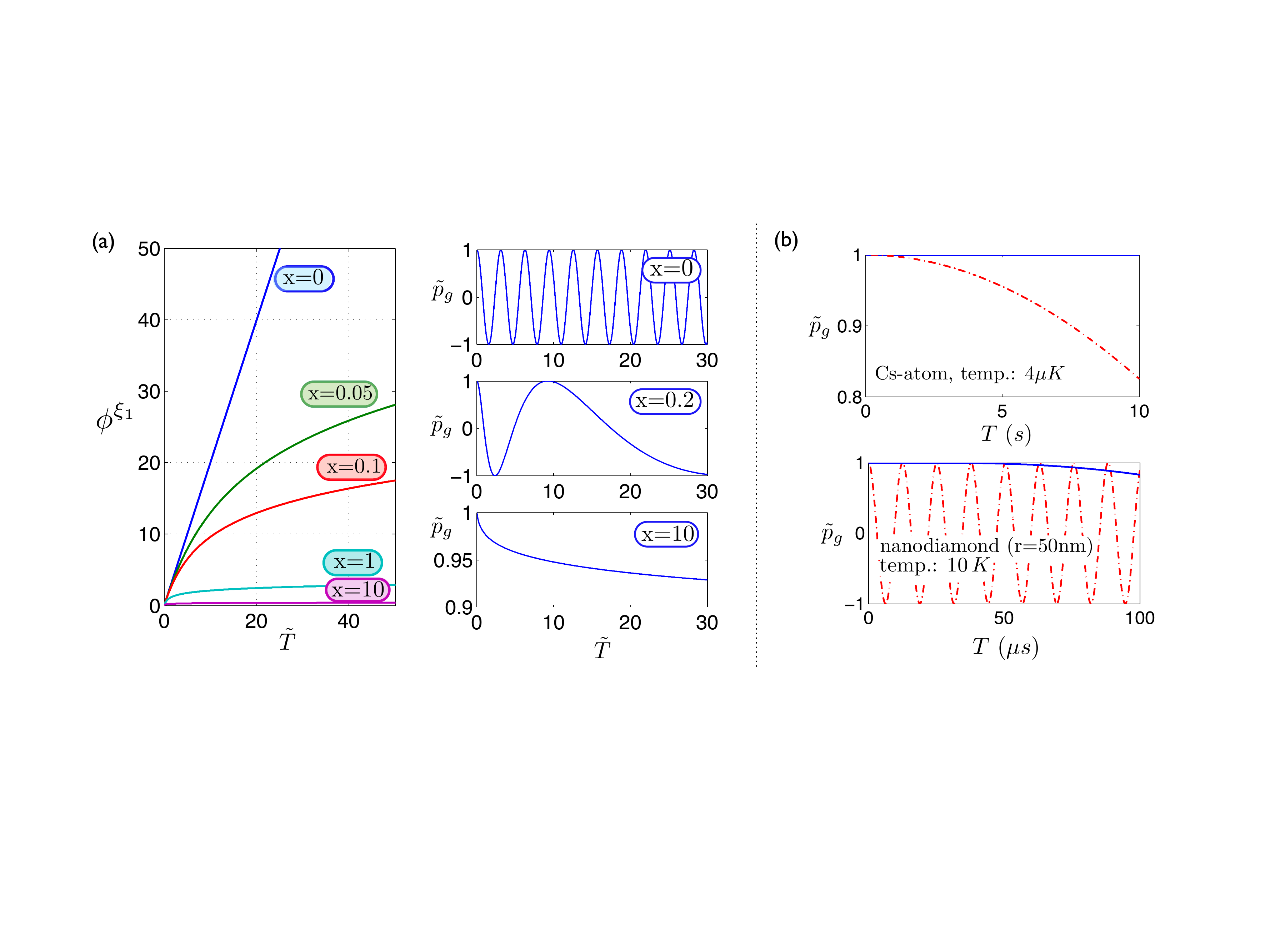}
\caption{\label{b_gravsup} (Color online) \textbf{Gravitational suppression and combined simulations.} \textbf{(a)} Influence of gravitation on the $\xi_1$-phase $\phi^{\xi_1}$ and oscillation fringe suppression $p_g'=1/8\,(1+\tilde{p}_g)$ for different values of $x=M_p\,g/(\hbar\,k^2 c)$. A zero initial momentum is assumed and the gravitational direction has been chosen orthogonal to the recoil momentum.  \textbf{(b)} $\xi_1$-simulations for realistic parameters combining gravitation ($\vec{g}\perp\vec{k}$) and thermal effects for Cs and a nanodiamond with radius 50\,nm \textit{(blue solid lines)} and the perfect unperturbed situation for comparison \textit{(red dashed-dotted lines)}. For Cs gravitation dominates whereas in the nanodiamond case both thermal and gravitational effects are of the same magnitude. This latter situation leads to an enhanced decoherence compared to the value given in table\,\ref{t_sigval} as is explained in the main text.  }
\end{centering}
\end{figure*}

The behaviour of the interference term in each of the regimes for a three-dimensional thermal momentum distribution is illustrated in figure\,\ref{b_thermaldecay} and exhibits exactly the previously discussed behaviour. 
Only the regime $\tilde{\sigma}\ll 1$ is appropriate to analyze the $\xi_1$-phase as expected and there is a clear significance of the previously described regimes. We note that in the overdamped regime a coherent oscillation on the $T_2$-timescale emerges, representing the very small non-zero average frequency in that regime.  The coherence times $T_2$ are  smallest in the $\tilde{\sigma}\sim 1$ regime and increase linearly with $\tilde{\sigma}$ and $1/\tilde{\sigma}$ in the $\tilde{\sigma}\gg1$ and $\tilde{\sigma}\ll 1$ limit, respectively. Importantly, whereas the relevant regime depends only on the value of $\tilde{\sigma}\propto \sqrt{m}$, the absolute timescales are proportional to $m^{-1}$ as expected by the mass dependent prefactor $\mu$ in (\ref{xi2}).  Additional non-zero average particle momenta $\ex{p_i}=p_0\neq{0}$ will push the interferometer further into the $\tilde{\sigma}\gg 1$ regime as is readily seen from figure\,\ref{b_phasevsp}\,(a), and therefore into a regime where neither a coherent nor a decoherent signature of the $\xi_1$-term will be observed, which then might lead to the premature conclusion that such a term does not exist. 

Typical values for the momentum variance $\tilde{\sigma}$ and the coherence time $T_2$ are shown in table\,\ref{t_sigval}. Originating from the small recoil momentum transfer those values are in general located far in the overdamped $\tilde{\sigma}\gg1$ regime, except for challenging low temperatures. As a consequence of the variance scaling with the particle mass, reaching the regime of coherent oscillations ($\tilde{\sigma}\ll 1$) seems very hard for more massive particles and might at most be possible for small nanodiamond radii ($\sim 5\,{\rm nm}$), following a harmonic ground state cooling with subsequent adiabatic relaxation of the trap frequency. Note that increasing the beam-splitter momentum $\hbar k$\,\cite{chiow09, chiow11, bouchendira11}, recalling that $\tilde{\sigma}=\sigma/(\hbar k)$, can help in the achievement of a coherent regime.

\subsubsection{Influence of gravitation}\label{sect_gravit}
For phase contributions quadratic in the momentum, gravitation leads at most to an additional phase factor (see also Appendix\,\ref{append_1}).  In the case of the $\xi_1$-contribution however, it can have a destructive effect on the observation of the phase itself, as follows out of figure\,\ref{b_phasevsp}\,(a).  With no additional pulse sequences involved, the gravitational effect can be described by choosing $\vec{p}_1(t')=\vec{p}_2(t')=\vec{p}^{\,0}-m\,\vec{g}\,t'$ in (\ref{xi2}), with a potential non-zero initial momentum $\vec{p}^{\,0}$, that we will set to zero for now. Such a situation is characterized by the dimensionless quantity $x=\left[mg/(\hbar k)\right]/\left[ 2\mu\hbar k \right]=M_p\,g/(\hbar\,k^2\,c)$, which corresponds to the ratio between suppression and coherent (optimal) phase evolution frequency. As both quantities scale proportional to the mass, this regime characterizing factor is mass independent, i.e. there exists no advantage concerning the fringe observation by changing the mass. For $x>1$ the overdamped regime is reached, in which coherent oscillations are essentially suppressed. In that regime ($mgT\gg \hbar k$) the phase\,(\ref{xi2}) can be approximated by $\phi_{\xi_1}\simeq 1/(2x)\,[1/2+\log(4x\tilde{T}) ] $
with $\tilde{T}=2\mu\hbar k\,T$.
In contrast $x\ll 1$ forms the desired regime in which gravitation is merely a perturbative factor to the $\xi_1$-phase evolution. However for a Raman recoil contribution with wavelength $\lambda=637\,{\rm nm}$, $x=1.7\cdot 10^4$, a value located deep in the suppression regime independent of the particle's size. 
A challenging recoil of $\sim 10^2\hbar k$ would be required to turn this into a coherent oscillatory $x\ll1$ regime, potentially feasible with atomic systems\,\cite{chiow09, chiow11, bouchendira11}. Figure\,\ref{b_gravsup}\,(a) illustrates this phase behaviour for different $x$-values showing a slowdown of the accumulated phase and consequently a fringe suppression for increasing magnitudes. For realistic $x$-values, no oscillations would be observable at all.

\subsubsection{Combining both effects}
Up to now, the thermal and gravitational influences have been analyzed separately. The combination of both is illustrated in figure\,\ref{b_gravsup}\,(b), for the case of a Cs atom and a nanodiamond under realistic experimental temperatures. As expected from the limiting cases discussed in the two preceding sections, $\xi_1$-oscillations are highly suppressed.  In cases where $\chi_{\rm grav}\equiv 2\,m\,g\,T\gg \sigma$, i.e. when the gravitational momentum gain is large compared to the typical thermal momentum magnitudes, the situation is well described by the pure gravitational case as is the case for the Cs atom simulation ($\tilde{\chi}_{\rm grav}=\chi_{\rm grav}/(\hbar k)\sim2\cdot 10^4$, $\tilde{\sigma}=1.7$). In the opposite regime $\chi_{\rm grav}\ll\sigma$ the thermal description holds. However, in an intermediate regime of comparable magnitudes the behaviour can deviate from the individual ones. Such a regime allows for the observation of $\xi_1$-correction signatures in the interference signal.  This is best understood in the (relevant) $|p|\gg|\hbar k|$ regime approximated in\,(\ref{xi3}). For a significant phase the first non-decaying term, reflecting the parallel component, has to be non-zero. This is only the case if there occurs a momentum angle change with respect to the recoil momentum axis in between the first and second interferometric cycle. For a purely thermal or gravitational situation this angle remains fixed. However combining both, gravitation will induce a change in the relative $p_z$ component, leading to $\theta_1\neq \theta_2$ for any thermal $\vec{p}$ non-parallel to the gravitational direction. This leads to a significant phase contribution in the limit when both the thermal momentum and the gravitational momentum gain are of the same magnitude.  In the absence of any initial effective momentum this does lead to a fluctuating contribution ($\ex{\cos\theta_1-\cos\theta_2}=0$), resulting in an increased decoherence instead of a coherent contribution, an effect that can be seen in the nanodiamond simulation of figure\,\ref{b_gravsup}\,(b). In that case $\tilde{\chi}_{\rm grav}\simeq 2\cdot 10^6$ $\sim \tilde{\sigma}=8\cdot 10^6$; for a purely thermal influence a much smaller decay rate would have to be expected from table\,\ref{t_sigval}.  Note however, that a non-zero initial momentum combined with gravitation can lead to a significant real coherent evolution based on the same enhancement mechanism, a discussion that will be the subject of the next section.
  
In summary, only two situation can lead to the observation of a significant coherent $\xi_1$-phase contribution: The first is cooling of the particle to a regime $\tilde{\sigma}\ll1$, requiring ground state cooling for trap frequencies $<1{\rm Hz}$ in the typical nanodiamond regime, along with performing the setup in a zero-gravity space. Both of those conditions could be significantly relaxed by increasing the recoil transfer of the beam splitter operation\,\cite{chiow09, chiow11} by at least a factor of $10^2$. Second, preparing the system in a non-zero initial momentum state non-parallel to the gravitation direction, with both the gravitational momentum transfer and the initial momentum being of the same magnitude and exceeding the momentum variance.

\section{Stability configuration for the $\xi_1$-measurement, resilient to gravitation and initial (thermal) momenta }\label{sect_stab}
\begin{figure}[htb]
\begin{centering}
\includegraphics[scale=0.42]{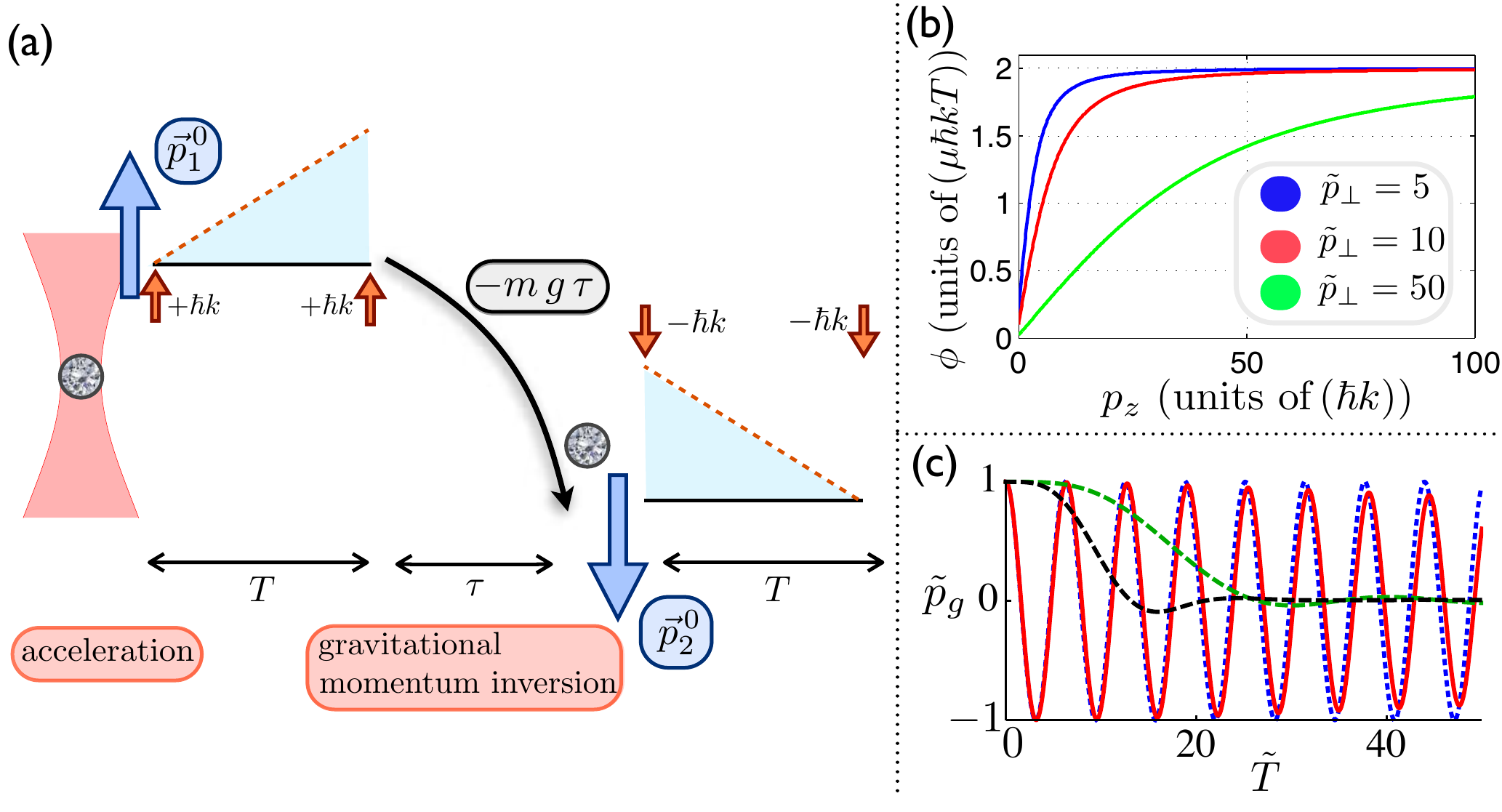}
\caption{\label{b_recovery1} (Color online) \textbf{Stability regime setup and phase recovery.} \textbf{(a)} Interferometric setup in the stability $\xi_1$-regime using gravitation for the momentum inversion.  \textbf{(b)} $\xi_1$-phase for different values of the recoil orthogonal momentum $\tilde{p}_\perp=p_\perp/(\hbar k)$, leading to a suppression of the phase term, vs. an increasing parallel component $p_{1,z}^{\,0}=-p_{2,z}^{\,0}=p_z$. For $|p_z|\gg |p_\perp|$ the stability regime is reached leading to the maximal optimal phase $\phi_{\xi_1}=2\mu\,\hbar k T$.  \textbf{(c)} Recovery configuration for $\tilde{\sigma}=\sigma/(\hbar k)=100$ and $x=M_p\,g/(\hbar k^2 c)=5$ in a configuration as depicted in (a). $\tilde{T}$ denotes the dimensionless time $\tilde{T}=2\mu\hbar k T$. Herein $\tilde{p}_z^{0}=p_z^{0}/(\hbar k)=x\,\tilde{T}+(1+\zeta\,\sqrt{2})\,\tilde{\sigma}$ with $\zeta=5$ guaranteeing that the parallel component exceeds the perpendicular one most likely by a factor of $\zeta=5$ even at its smallest value at time T (assuming equal momentum variances $\tilde{\sigma}$ in all spatial directions). The time $\tau$ is chosen to allow for the same ratio in the second cycle, such that $p_{2,z}^{\,0}\simeq -\zeta \sqrt{2}\sigma$. Red solid lines show the $\xi_1$ population oscillations in the recovery regime, blue dotted lines the optimal phase without any disturbance, whereas green (upper) and black (lower) dashed lines correspond to a non-adjusted `normal' configuration with $\tau=0$ and $\vec{p}_1^{\,0}=0$ for $\vec{g}\perp\vec{k}$ and $\vec{g}\parallel \vec{k}$, respectively.   }
\end{centering}
\end{figure}
As outlined in the previous section, a change in the momentum direction relative to the recoil orientation $\vec{k}$ within the interferometric sequence, can lead to the appearance of a significant non-zero phase term even in the presence of a large absolute momentum value. In the following the terms `parallel' and `orthogonal' will always refer to the momentum relative to the recoil direction $\vec{k}$ of the first interferometric cycle $[t_0,t_1]$, unless stated otherwise. Whereas for the orthogonal component each individual cycle phase contribution decays to zero with increasing $|p|$, for the parallel component the combination of both cycles is crucial. Changing the direction of the initial momenta, such that $\vec{p}_1^{\,0}\neq\vec{p}_2^{\,0}$ (see figure\,\ref{b_recovery1}), or more precise changing the angle to the recoil term, will lead to a momentum independent contribution (see (\ref{xi3})). This fact can be used to construct a regime, in which the $\xi_1$-phase can be observed despite thermal and gravitational influences.   
For static momenta, the optimal condition can be identified from\,(\ref{xi3}), following as
\begin{equation}  \vec{p}_1^{\,0}\uparrow\uparrow \vec{k} \quad \&\&\quad \vec{p}_2^{\,0}\uparrow\downarrow \vec{k}\,\,(\Leftrightarrow  \vec{p}_1^{\,0} \uparrow\downarrow \vec{p}_2^{\,0})   \end{equation}
i.e. the first initial momentum is parallel to the recoil whereas there occurs a change in the momentum direction in between the two interferometer cycles, such that the second initial momentum is antiparallel to the first one (and parallel to the second recoil direction). In that case $\phi_{\xi_1}=\mu\,T\,2\,\hbar k$, which corresponds to the optimal phase (\ref{bound1}), the same as would appear in the absence of any momenta except the beam splitter operations. We will call this regime the `stability regime', because, once the parallel and antiparallel condition is reached, the phase is highly stable and independent of the absolute momentum value, thereby  making it stable against decoherence. Moreover this holds for any momentum even outside the $|p|\gg|\hbar\,k|$ regime as will be shown in Appendix\,B.  One could anticipate from\,(\ref{xi3}) that a configuration $\vec{p}_1^{\,0}\uparrow\downarrow \vec{k}$ and $\vec{p}_1^{\,0} \uparrow\downarrow \vec{p}_2^{\,0}$ will lead to the same phase magnitude despite a negative overall sign, however such a regime is only momentum independent in the limiting case of the validity of\,(\ref{xi3}). A configuration based on changing the parallel momentum direction in between the first and second cycle, is demonstrated in figure\,\ref{b_recovery1}\,(b). Once the parallel regime is approached, i.e. the parallel momentum dominates the orthogonal component, the phase is recovered up to its optimal unperturbed value.

\begin{figure}[htb]
\begin{centering}
\includegraphics[scale=0.4]{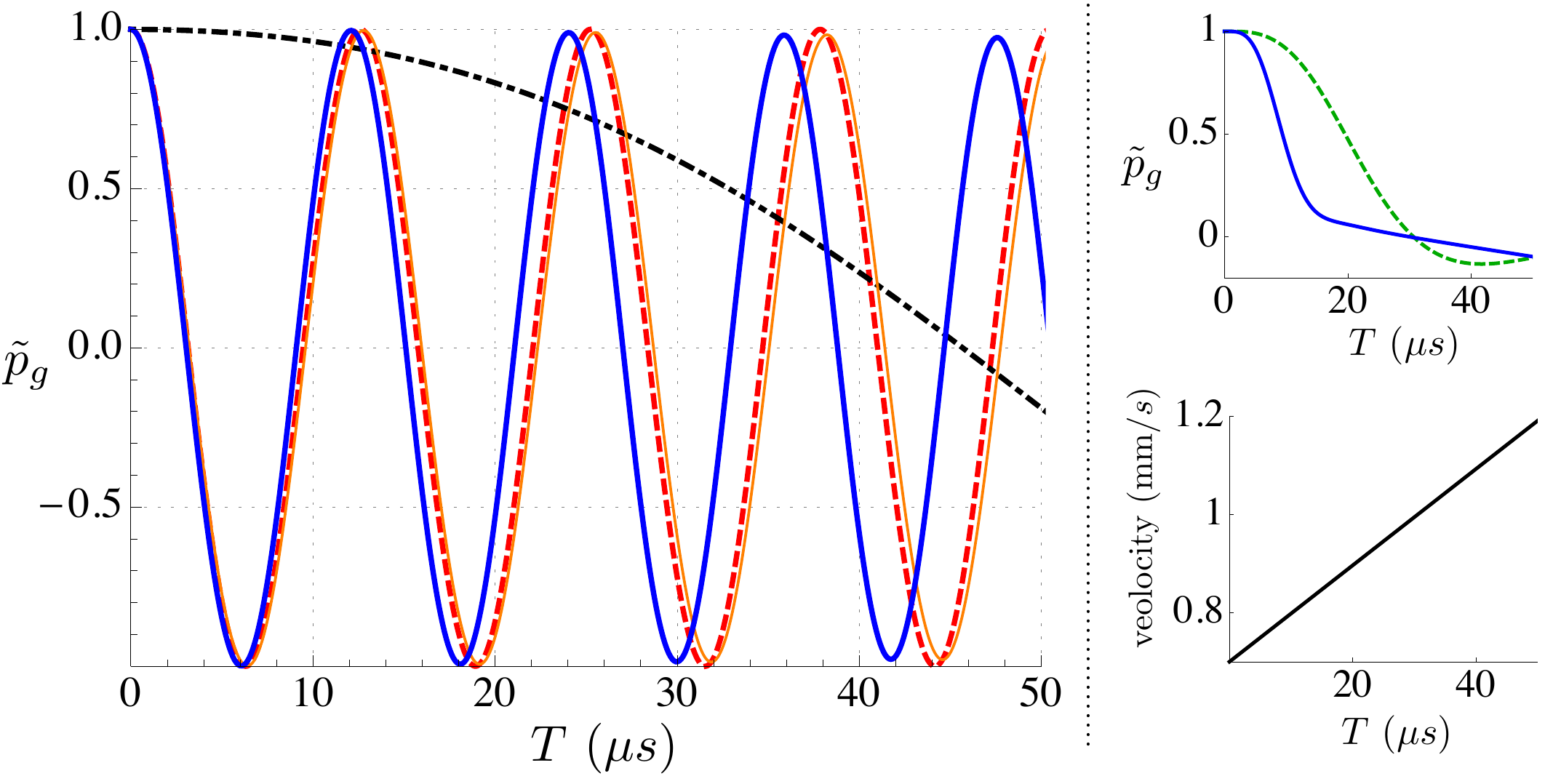}
\caption{\label{b_recovery2} (Color online) \textbf{Phase recovery for a 50\,nm nanodiamond.} Phase recovery scheme simulation for a setup as depicted in figure\,\ref{b_recovery1}\,(a), a nanodiamond of radius 50\,nm, and a parallel component exceeding the orthogonal one upon design by a factor $\zeta=5$ as defined in the caption of figure\,\ref{b_recovery1}. The temperature has been chosen as $1\,{\rm mK}$ in order to make the thermal momentum variance comparable to the gravitational momentum gain. \textit{Left:} Population oscillations  for the combination of zero order and $\xi_1$-term \textit{(blue solid)}, the $\xi_1$-term alone \textit{(red dashed)} and the zero order contribution alone \textit{(black dashed-dotted)}.  The orange (\textit{thin solid}) line indicates the optimal $\xi_1$ oscillations in the absence of thermal noise and gravitation. \textit{Upper right: } Interference oscillations in the absence of a specifically designed recovery sequence for the same parameters, but $\tau=0$ and $p_1^{\,0}=0$ and $\vec{g}\parallel \vec{k}$ (\textit{blue solid}) and $\vec{g}\perp\vec{k}$ (\textit{green dashed}), respectively.  \textit{Lower right:} Initially transferred velocity parallel to the recoil vs $T$.   }
\end{centering}
\end{figure}

Experimentally, one has merely to guarantee the parallel condition along with a momentum direction change in between the two cycles involved. The first one can be arranged by an initial parallel momentum sufficiently large such that it exceeds the momentum gain of gravitation and the thermal momentum variance $\sigma$. The second condition  will require a directional momentum change during the intermediate time period $\tau$, that can be obtained by an external acceleration or making use of gravitation. We will focus on the latter, as a momentum transfer by e.g. laser pulses may be prone to destroy the spatial superposition state by e.g. incoherent photon scattering. This leads to a possible setup as depicted in figure\,\ref{b_recovery1}\,(a) in a configuration such that $\vec{g}$ is antiparallel to the initial recoil direction. An initial momentum is transferred  such that its magnitude exceeds the thermal variance and gravitation, $|\vec{p}_1^{\,0}|>\sigma, 1/2\,\chi_{\rm grav}$, ensuring the parallel regime, followed by a free evolution sequence $\tau$ in which gravitation provides the change to an antiparallel configuration with $|\vec{p}_2^{\,0}|>\sigma$. For such a gravitational scheme to work, the thermal variance should be at most comparable (or smaller) in magnitude to the gravitational momentum gain $\sigma\lesssim|\chi_{\rm grav}|$  as otherwise the intermediate time period would have to be much longer than the interferometer time ($\tau\gg T$). This will in general imply an initial motional particle cooling\,\cite{kiesel13} more restrictive for increasing particle mass. Figure\,\ref{b_recovery1}\,(c) demonstrates that mechanism in dimensionless units. For a parallel momentum component exceeding the orthogonal fluctuating thermal ones by at least a factor of five, an almost perfect recovery of coherent oscillation fringes is achieved. A realistic example is provided in figure\,\ref{b_recovery2} for a nanodiamond of radius 50\,nm. Here an initial cooling to a temperature of $1{\rm mK}$ along with an initial velocity transfer of the order of 1\,mm/s, allows for the observation of well-defined coherent interference fringes on a timescale of $200\,\mu s$, within range of the NV-center coherence time. This would be a sufficient test for the existence of such a term.

Note that, by working in the parallel regime, in particular in a regime with the gravitation being parallel to the recoil direction, the zero order phase contribution will be altered as well, namely by an additional phase $\Delta\phi_0=(1/m)\,\vec{k}\,\left[  \vec{p}_1^{\,0}-\vec{p}_2^{\,0} \right]\,T$. This term will have to be accounted for in a proper determination of $\xi_1$, requiring a precise knowledge of the initial momentum and the gravitation direction.  To avoid that additional change one could also use alternative momentum schemes by choosing the gravitation orthogonal to $\vec{k}\parallel \vec{p}_1^{\,0}$, with the initial momentum direction subsequently changed by at most a $\pi/2$ rotation for the second cycle under the influence of gravitation; however, with the drawback that such a regime is only approximately momentum-independent  for $|p|\gg |\hbar k|$ and according to\,(\ref{xi3}) the maximal phase is one half of the one in the parallel gravitational approach.

\section{Total interference pattern: multi-path interference}\label{sect_totint}
Up to now we have considered a single closed interferometric path combination. Aside of demonstrating the basic concepts such a description holds whenever the spatial separation to neighbouring paths is large, such that additional interference can be neglected and only a single path lies within the detection region. This is generally fulfilled  for interferometric setups with atoms, where the single path model is widely used\,\cite{wicht02, cadoret09, bouchendira11}. For an implementation with nanodiamonds however, noting that the spatial separation $\propto (\hbar k/m)\,T$ scales inversely with the mass, the path splitting turns out to be very small. As a consequence, the influence of additional paths will have to be taken into account. 
 
\begin{figure*}[htb]
\begin{centering}
\includegraphics[scale=0.5]{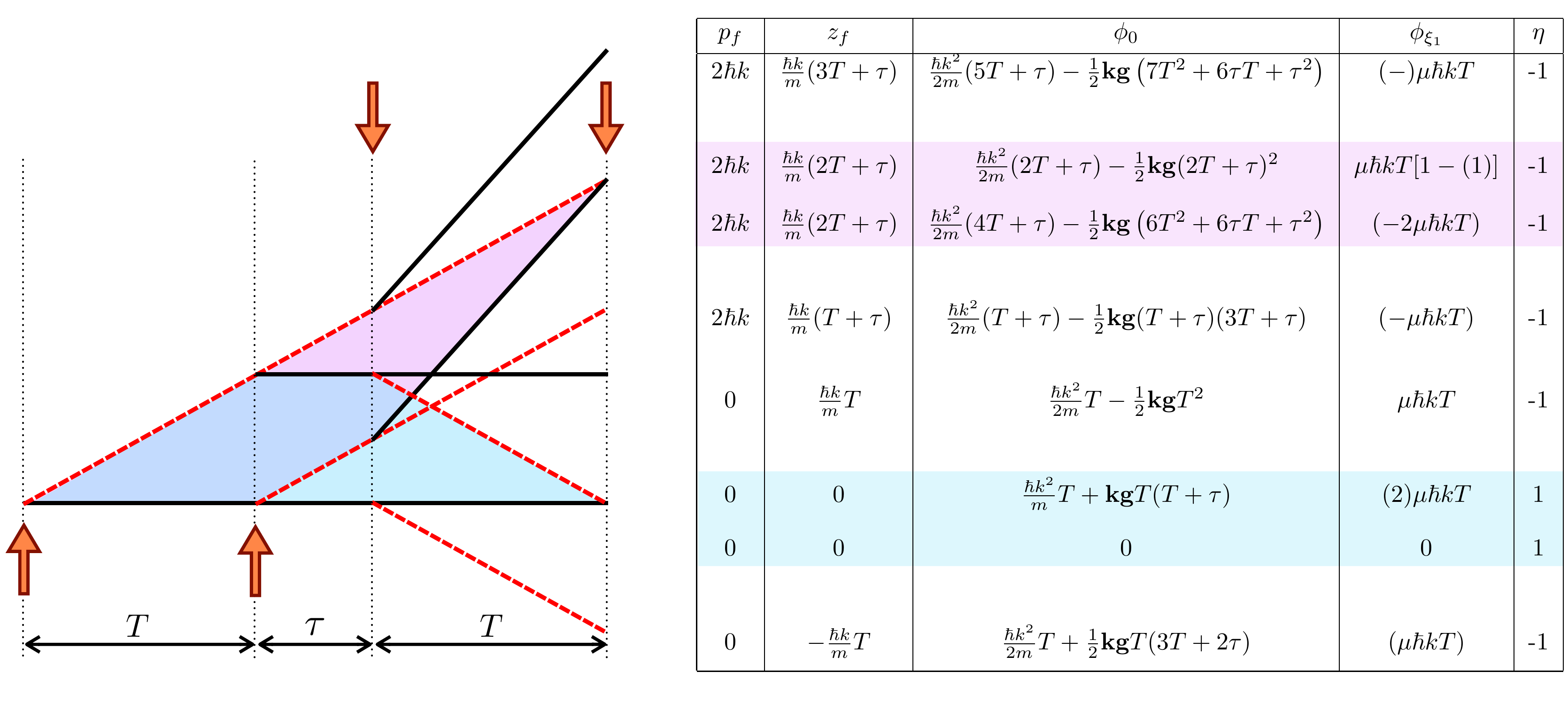}
\caption{\label{b_allpath} (Color online) \textbf{Ramsey-Bord\'e paths and interference phase contributions.} \textit{Left:} The complete path scheme involved in the interferometer sequence with red arrows denoting Raman-$\pi$/2 pulses and black and red (dashed) lines corresponding to the particle internal state being in the ground and excited state, respectively. The two closed interferometer configurations are highlighted by the cyan and pink shaded areas. \textit{Right:} Path contributions to the interferometer phase. $p_f$ denotes the total momentum gain, the difference for different paths directly corresponding to the path separation in momentum $\Delta p$, and $z_f$ the final position (both neglecting accelerations that are common to all paths and therefore will have no impact on the interference). $\phi_0$ describes the momentum independent phase part relative to the second path ($\phi_2(p)$) from below, and $\phi_{\xi_1}$ the quantum gravity phase part ($\mu=\xi_1m c/(2\hbar M_p)$). For the latter, terms in curved brackets are absent (present) in a setup with the gravitation orthogonal (parallel) to the recoil direction $\hbar \mathbf{k}$. $\eta$ accounts for the beam-splitter phase as defined in (\ref{totint1}) \&  (\ref{totint3}).  The total phase (\ref{totint5}) follows as $\phi_j(p)-\phi_2(p)=\phi_0^j+(1/\hbar) z_f^j\,p+\phi_{\xi_1}^j$ with $\phi_2(p)=1/(6m^2\hbar g)\,(p^3-[p-mg (2T+\tau)]^3)$.
A direct connection to the interference element form (\ref{totint13}) and (\ref{totint14}) is given by $\Delta p_{ij}=p_f^i-p_f^j$, $\Delta z_0^{ij}=z_f^i-z_f^j$ and $\Delta\phi_{ij}(0,\Delta p)=\Delta\phi_{ij}(0,0)-\Delta p/(2\hbar)(z_f^i+z_f^j-g[2T+\tau]^2)$ with $\Delta\phi_{ij}(0,0)$ the corresponding phase difference out of $\phi_0$ and $\phi_{\xi_1}$. }
\end{centering}
\end{figure*}

The total interference pattern is then determined by eight paths involved in the Ramsey-Bord\'e setup as illustrated in figure\,\ref{b_allpath}.  Importantly, a second closed path interferometric combination with equal area exists on top of the previously discussed partial interferometer. Two possible effects might arise by including the additional paths: First, the appearance of additional interference contributions, namely each path may interfere with each of the other paths. This then leads to additional phase terms and possibly additional frequencies in the interference pattern.  Second, an addition of the partial interference pattern from distinct spatial regions, that is an addition of populations as a consequence of the inability to resolve individual path combinations in the detection process. This second property even holds if these regions are independent in terms of interference.  Importantly, the ability for interference decreases with increasing separations in momentum or position of the pathways under consideration. Whereas interference of closed paths can be accomplished even with thermal (incoherent) particle states, for open paths this property depends crucially on the state coherence in momentum and position. Generally this coherence, and thus the interference contribution of open path combinations, decreases with increasing temperature. This behaviour manifests the predominant influence of the closed interferometer pathways, whose interference contribution corresponds to the one analyzed in the preceding sections. For sufficiently large temperatures, these pathways will in fact be the only significant ones contributing to the interference phase.

\subsection{Total interference pattern and interference fringe visibility decay}
The total unitary evolution, describing paths that do end up in the ground state $\ket{g}$ of the `two-level' particle, takes the form
\begin{equation}\label{totint1}   U_{\rm tot}^{(g)}=\left( \frac{1}{\sqrt{2}} \right)^n\,\sum_{i=1}^{2^{n-1}}\,\eta_i^{(g)} \,U_i^{(g)}  \end{equation}
with $n$ the number of $\pi/2$-pulses involved in the sequence (here $n=4$) and $\eta_i^{(g)}=(-i)^k$ with $k$ the number of population inversions of the corresponding path accounting for the beam splitter operation as defined in\,(\ref{ip1}).  $U_i^{(g)}$ describes the unitary evolution of path $i$, whose calculation is exemplified for the lower closed paths interferometer in Appendix\,\ref{append_1}. The evolution $U_{\rm tot}^{(e)}$ of all paths ending up in the internal excited state $\ket{e}$ and again leading to $2^{n-1}$ paths, can be described analogously by merely replacing the index $(g)$ by $(e)$ in\,(\ref{totint1}). Thus, the final state after the interferometric sequence follows as $\rho_{f}=U_{\rm tot}\,\rho_{\rm in}\, U_{\rm tot}^\dagger$ from the initial state $\rho_{\rm in}$ with $U_{\rm tot}=U_{\rm tot}^{(g)}+U_{\rm tot}^{(e)}$. The probability for finding the internal state in its ground state configuration is therefore given by
\begin{equation} p_g=\tr\left(\rho_f\,\ket{g}\bra{g}  \right)=\tr\left(U_{\rm tot}^{(g)}\,\rho_i\,U_{\rm tot}^{(g)\,\dagger} \right)\,.   \end{equation}
With the help of (\ref{totint1}) this can be re-expressed as
\begin{equation}\label{totint3}\begin{split} &p_g=\left(\frac{1}{2}\right)^n\,\tr\Biggl(\sum_{i,j=1}^{2^{n-1}} \left(\eta_i^{(g)} \eta_j^{(g)*}\right) U_i^{(g)}\rho_{\rm in}\, U_j^{(g)\,\dagger}  \Biggr)\\
&=\frac{1}{2}\Biggl(1+\frac{1}{2^{n-1}}\sum_{i<j}^{2^{n-1}}\left[ \left(\eta_i^{(g)} \eta_j^{(g)*}\right)\tr(U_i^{(g)}\rho_{\rm in}U_j^{(g)\dagger})+\text{h.c.}     \right] \Biggr)
  \end{split}\end{equation}
where in the last step the $2^{n-2}\,(2^{n-1}-1)$ interference terms have been separated from the pure population contributions. 
The evaluation of the total Ramsey Bord\'e setup ($n$=4) necessitates the calculation of 28 interference contributions, amongst two paths configurations closed  in position and momentum. \\Essentially this requires calculating contributions of the form $\tr(U_i^{(g)}\,\rho_{\rm in}U_j^{(g)\dagger})$, wherein the evolution operators can be expressed as (see Appendix\,\ref{append_1})
\begin{equation}\label{totint4}  U_j^{(g)}=\e^{(i/\hbar) p_{\rm f}^j\,\hat{z}}\,\e^{-i\phi_j(\hat{p})}  \end{equation}
with $p_f^j$ the total final (classical) momentum gain of the path $j$ through the beam splitter and accelerating forces during the interferometer sequence. Thus the first contribution in (\ref{totint4}) corresponds to a displacement operation in momentum space. The momentum dependent phase factor, by restricting to potentials at most linear in position, is shown in Appendix\,\ref{append_1} to equal the kinetic energy integrated along the classical interferometer path
\begin{equation}\label{totint5} \phi_j(\hat{p})=\frac{1}{\hbar}\,\int_{t_0}^{t_f} E(\hat{p}_j(t'))\,\mathrm{d}t'+\varphi_l^j \end{equation}
with $\hat{p}_j(t)$ the time dependent momentum in path $j$ and $t_0$ and $t_f$ the initial and final times, respectively. Herein, the momentum evolution operator corresponds to the classical momentum evolution with the initial momentum $p$  replaced by the operator analogue $\hat{p}$. The phase factor $\varphi_l^j$ accounts for the purely internal state dependent detuning and laser phase contributions and will be assumed to be zero in the following. Note that momentum operator dependent contributions of $\phi_j(\hat{p})$ in (\ref{totint4}) describe the path displacement in position space. Explicit expressions for the phase factor and the final momentum for the individual paths can be found in figure\,\ref{b_allpath}. 

With the general form of the unitary operators (\ref{totint4}) at hand, the interference terms evaluated in momentum space take the form
\begin{equation}\label{totint6}\tr\left(U_i^{(g)}\rho_{\rm in} U_j^{(g)\dagger}  \right)=\int\mathrm{d}p\, e^{-i\left[ \phi_i(\hat{p})-\phi_j(\hat{p}+\Delta p)  \right]} \ex{p|\rho_{\rm in}|p+\Delta p}    \end{equation}
with $\Delta p\equiv\Delta p_{ij}=p_{\rm f}^i-p_{\rm f}^j$ the final momentum difference of both paths. The evaluation in momentum representation is of particular advantage as it allows for the simple inclusion of modifications to the energy dispersion relation.  The calculation of (\ref{totint6}) involves the matrix element $\ex{p|\rho_{\rm in}|p+\Delta p}$, that characterizes the coherence (`momentum overlap') of the initial state in momentum space.  

This can be interpreted in that for starting in the two different momentum states both of the paths involved end up in the same final momentum state and therefore contribute constructively to the interference pattern. That is, an initial separation of the paths by $\Delta p$ in momentum space will translate into a perfect overlap in the detection region. Moreover the momentum distribution determined by the coherence element and the average (integration) involved accounts for the separation and overlap in position space.

We will calculate the matrix element in Appendix\,\ref{append_overlap} for an initial thermal harmonic oscillator state of frequency $\omega_t$, which leads to
\begin{equation}\label{totint7}  \ex{p|\rho_{\rm in}|p+\Delta p}= \mathcal{P}\left(p+\Delta p/2\right)  \,\e^{-\frac{1}{2\hbar^2}\Delta p^2\ex{\hat{z}^2}} \end{equation}
with $\mathcal{P}(p)$ a Gaussian distribution of variance $\ex{\hat{p}^2}$
\begin{equation}\label{totint8} \mathcal{P}(p)=\frac{1}{\sqrt{2\pi\ex{\hat{p}^2}}}\,\e^{-\frac{1}{2}\,p^2/\ex{\hat{p}^2}}  \end{equation}
and the variances given by
\begin{equation}\label{totint9} \ex{\hat{z}^2}=\frac{\hbar}{2m\omega_t}\,\left(2\ex{\hat{n}}+1  \right) \, \text{ and }\, \ex{\hat{p}^2}=\frac{m\omega_t\hbar}{2}\,\left(2\ex{\hat{n}}+1  \right)\,  \end{equation}
where the average phonon number $\ex{\hat{n}}=\exp(\hbar\omega_t/(k_B \mathcal{T})-1)^{-1}$. In the high temperature limit $k_B\mathcal{T}/(\hbar\omega_t)\gg1$ the momentum variance $\ex{\hat{p}^2}\simeq mk_B \mathcal{T}$ and (\ref{totint8}) reduces to the Boltzmann distribution independent of the trap frequency; in contrast the position variance $\ex{\hat{x}^2}\simeq k_B \mathcal{T}/(m\omega_t^2)$ retains its dependence on the initial localization (trap frequency).

Inserting (\ref{totint7}) and (\ref{totint8}) into (\ref{totint6}) allows then to rewrite the interference element as
\begin{equation}\label{totint10} \tr\left(U_i^{(g)}\rho_{\rm in} U_j^{(g)\dagger}  \right)=\e^{-\frac{1}{2\hbar^2}\Delta p^2\ex{\hat{z}^2}}\,\int \mathrm{d}p\,\mathcal{P}(p) \,\e^{-i\,\Delta\phi_{ij}(p,\Delta p)} \end{equation}
with the definition of the phase difference
\begin{equation} \Delta\phi_{ij}(p,\Delta p)=\phi_i(p-\Delta p/2)-\phi_j(p+\Delta p/2)\,.   \end{equation}
That way, based on the momentum difference of the paths involved and their phase\,(\ref{totint5}), the interference matrix element (\ref{totint10}) can be readily evaluated. Whereas the decay due to a path separation in momentum appears directly in expression\,(\ref{totint10}), the decay due to a separation in position as characterized by  the momentum dependent terms in $\Delta \phi$  follows as a result of the momentum averaging process. 

In the following we will make this decay behaviour more explicit by assuming that $\Delta \phi_{ij}$ can be rewritten in the particular form
\begin{equation}\label{totint12}  \Delta\phi_{ij}(p,\Delta p)=\Delta\phi_{ij}(0,\Delta p)+(1/\hbar)\,\Delta z\,p\,,  \end{equation}
which holds for the standard form of the energy dispersion relation $E(p)=p^2/(2m)$ as well as in the `stability configurations'; in the latter case the quantum gravity phase part is momentum independent, or in other words does not lead to additional (final) path separations in position. Herein $\Delta\phi_{ij}(0,\Delta p)$ corresponds to a pure phase whereas $\Delta z$, recalling its origin from a displacement operator in (\ref{totint4}) evaluated in momentum space, describes the path separation in position.  That way (\ref{totint10}) can be calculated to
\begin{equation}\label{totint13}  \tr\left(U_i^{(g)}\rho_{\rm in} U_j^{(g)\dagger}  \right)=\e^{-i\Delta\phi_{ij}(0,\Delta p)}\e^{-\frac{1}{2\hbar^2}\Delta p^2\ex{\hat{z}^2}} \e^{-\frac{1}{2\hbar^2}\Delta z^2\ex{\hat{p}^2}}  \end{equation}
which does have an intuitive interpretation in that path separations in position and momentum lead to a decay of the interference element with the momentum and position variance of the initial thermal state, respectively.  This decay is more pronounced for an increasing temperature, which reduces the state coherence. $\Delta\phi_{ij}(0,\Delta p)$ takes the role of the relevant phase factor. 
It is worth noting that, whereas $\Delta p$ corresponds to the classical momentum difference of the paths involved, the analogue interpretation for $\Delta z$ holds true only in cases of $\Delta p=0$. Otherwise, as the interference is then characterized by the two contributing paths starting in different momentum states as indicated by the overlap matrix element of (\ref{totint6}), the effective separation in position follows as
\begin{equation} \label{totint14} \Delta z=\Delta z_0-\frac{\Delta p}{m}\,t_{\rm tot}\,.   \end{equation} 
Here $\Delta z_0=\hbar (\mathrm{d}/\mathrm{d}p)\Delta\phi_{ij}(p,0)$ denotes the classical path difference for a particle with a well defined initial momentum and the second contribution accounts for the deviation from that situation, namely initial momenta that differ by $\Delta p$. Remarkably, this means that closed paths in position but differing in momentum, can be effectively open in position in the evaluation of (\ref{totint13}); though this is not the case in the interferometric setup under consideration.

\subsection{Nanodiamond interference setup including all paths}
For nanodiamonds, the final path separation $\Delta z\propto 1/m$ turns out to be very small, which prevents the selection of individual paths in the detection process and in addition may lead to interference contributions of additional paths other than the closed ones. As can be seen in figure\,\ref{b_allpath}, the typical path separations in position and momentum are given by $\Delta z\simeq (\hbar k/m) T\sim 1/m$, or $\sim 1/m^2$ on the typical period time $T'=2\pi/(2\mu\hbar k)\sim 1/m$ of the interference fringes, and $\Delta p=2\hbar k$, respectively. In contrast, the state overlap in position and momentum space as defined in (\ref{totint13}) together with (\ref{totint9}) follows as
\begin{equation}\label{totint15}  \Delta z_r=\sqrt{\frac{4\hbar}{m\omega_t\,[2\ex{\hat{n}}+1]}}, \quad \Delta p_r=\sqrt{\frac{4\hbar m\omega_t}{2\ex{\hat{n}}+1}}  \end{equation}
leading to a visibility reduction $\mathcal{V}_{\Delta z}=\exp(-\Delta z^2/\Delta z_r^2)$ and $\mathcal{V}_{\Delta p}=\exp(-\Delta p^2/\Delta p_r^2)$. Therefore, the interference of open paths in measuring a possible $\xi_1$-quantum gravity contribution becomes more probable with increasing particle mass. 

It is worth considering first the impact of the two closed paths interferometric configurations. Their contribution to the interference pattern is independent of the initial particle state, thus independent of temperature and the state coherence overlap as defined in (\ref{totint15}). Thus, for any particle mass the final population takes the form
\begin{equation}\label{totint16}  p_g^{\rm closed}= \frac{1}{2}\,\left(1+\frac{1}{4}\,\left[ \cos\phi^{l}+\cos\phi^{u} \right] \right) \end{equation}
with $\phi^l$ and $\phi^u$ the phase of the upper and lower closed interferometric combination ($\mu=\xi_1m c/(2\hbar M_p)$)
\begin{equation}\label{totint17} \phi^{l/u}=\pm\frac{\hbar k^2}{m}\,T+\vec{k}\,\vec{g}\,T(T+\tau)+2\mu\hbar kT\,.  \end{equation}
For a general momentum inversion the gravitational term can be replaced by $\vec{k}[\vec{p}_1^{\,0} -\vec{p}_2^{\,0}] (T/m)$  with $\vec{p}_1^{\,0}$ and $\vec{p}_2^{\,0}$ the initial momenta and the one at time $T+\tau$, respectively. As can be seen out of figure\,\ref{b_contribscale}, the recoil contribution ($\hbar k^2 T/m$) is significantly smaller for nanodiamonds compared to the anticipated quantum gravitational corrections such that $\phi^l\simeq \phi^u$. Thus the constructive addition of the two closed paths combinations leads to a doubling of the interference fringe contrast. In case that interference of open pathways contributes, this will lead to additional phase terms in (\ref{totint16}).

Noting that the $\xi_1$-phase contribution increases linearly with mass, larger diamonds of radius $\sim 50\,{\rm nm}$ have been identified optimal in section\,\ref{sect_stab} for observing $\xi_1$-induced interference fringes within the internal state coherence time. As stated at the beginning of this section, this makes the interference of open path combinations more probable.

\begin{figure}[thb]
\begin{centering}
\includegraphics[scale=0.5]{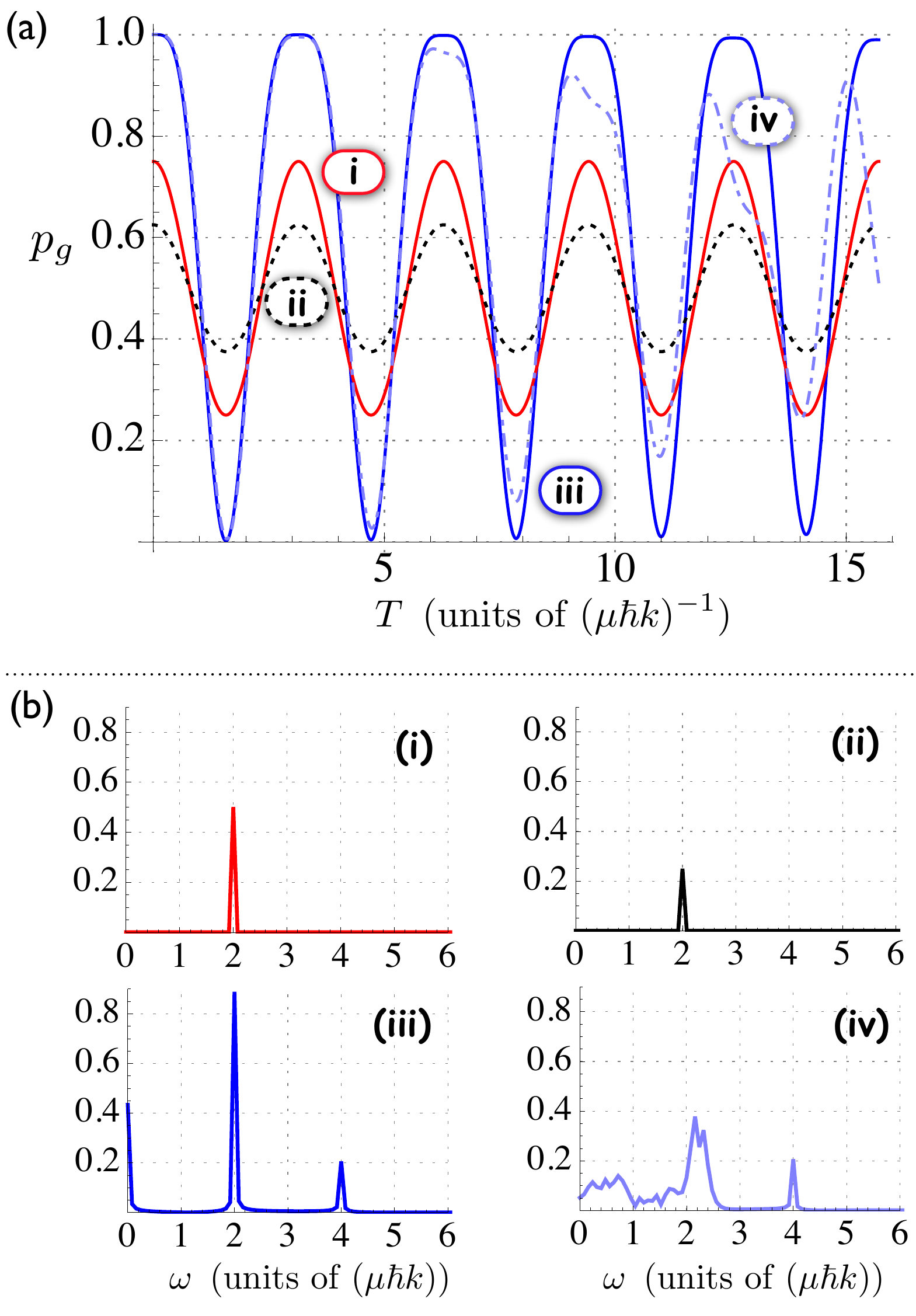}
\caption{\label{b_allpath_interf} (Color online) \textbf{Interference pattern comprising all paths. } \textbf{(a)} Ground state population after passing the Ramsey-Bord\'e interferometer vs interferometer time $T$ assuming a `stability configuration' and a linear QG correction to the energy dispersion relation as defined in\,(\ref{pqg1}). (i)-(iii) correspond to the QG phase part alone for (i) (\textit{solid red}) both closed interferometric path combinations ($r=10\,{\rm nm}$, $\mathcal{T}$=1\,mK, $g$=0) and (ii) (\textit{black dashed}) a single closed path interferometer contribution and (iii) (\textit{solid blue}) all paths of the interferometer with an (almost) perfect state overlap guaranteeing the interference of all paths involved ($r=50\,{\rm nm}$, $\mathcal{T}$=1\,mK, $g$=0). (iv) (\textit{blue dashed dotted}) represents a r=50\,nm diamond at a temperature of $1\,{\rm mK}$ for the total phase including gravitation. \textbf{(b)} Frequency contributions of the interference oscillations for the situations as illustrated in (a) based on a FFT analysis over 25 oscillation periods. For $\xi_1\simeq 1$, $\mu\hbar k\simeq 2\pi\cdot 41\,{\rm kHz}$ for a diamond of radius 50\,nm ($\mu=\xi_1m c/(2\hbar M_p)$).}
\end{centering}
\end{figure}

 Let us assume a nanodiamond of radius 50\,{\rm nm},  the initial localization to be characterized by a trap frequency $\omega_t=2\pi\cdot 165\,{\rm kHz}$, a temperature of $1\,{\rm mK}$ and an interferometer time of $T=500\,\mu s$. This corresponds to a configuration as in figure\,\ref{b_recovery2}. Paths open in position are then characterized by a typical path separation $\Delta z\simeq 0.6\,{\rm pm}$, leading to an interference term visibility of  $\mathcal{V}_{\Delta z}\simeq 0.7$, and $\mathcal{V}_{\Delta p}\simeq 1$. On the timescale of a single oscillation fringe ($T'=(2\pi)/(2\mu\hbar k)=12\,\mu s$), the path separation is even smaller and $\mathcal{V}_{\Delta z}\simeq 1$. That is, all path combinations contribute significantly to the interference pattern.\\
The expected interference pattern and the frequency contributions involved are illustrated in figure\,\ref{b_allpath_interf}; characterized in this configuration by an almost perfect coherence overlap and thus interference comprised of all paths involved and a full contrast in the fringe visibility. For clarity, the pure quantum gravity phase contribution is shown by configuration (iii) in figure\,\ref{b_allpath_interf}. It is characterized by the frequency $\omega_{\xi_1}= 2\mu\hbar k$, as already appears in the exclusive interference of the closed paths (\ref{totint17}), and additional frequency contributions at the double frequency $2\omega_{\xi_1}$ along with a static contribution.  Including the effect of gravitation, crucial in case it is used as the underlying mechanism for momentum inversion, results in an interference pattern as illustrated by (iv).    The gravitational influence, or the momentum inversion component, introduces an additional frequency modulation in time that tends to smear the sharp frequency characteristic. 

Decreasing the nanodiamond size, a viable scenario for larger internal coherence times, changes the expected interference pattern significantly. That is, the path separation increases and the interference pattern is gradually reduced towards the situation with interference terms exclusively arising from the closed paths contributions (\ref{totint16}) and (\ref{totint17}). For a diamond of radius 10\,nm and otherwise equal parameters ($\omega_t=2\pi\cdot 165\,{\rm kHz}$, $1\,{\rm mK}$, $T=500\,\mu s$), the visibility decay of spatially open paths is characterized by $\mathcal{V}_{\Delta z}\simeq 10^{-20}\simeq 0$ and $\mathcal{V}_{\Delta p}\simeq 0.97$. Thus interference contributions of open paths can be essentially excluded.
The interference pattern contribution for the $\xi_1$-phase then takes the form as illustrated by configuration (i) in figure\,\ref{b_allpath_interf}, characterized by a single frequency $\omega_{\xi_1}=2\mu\hbar k$ and a reduced fringe contrast. Note that, as the coherence overlaps\,(\ref{totint15}) decrease with temperature, such a reduction of the interference pattern to the closed paths contribution also holds true for larger particles by increasing the temperature.

The additional frequency modulation in time introduced by gravitation can be avoided by an orthogonal regime as described at the end of section\,\ref{sect_stab}, at the drawback of reducing the $\xi_1$-phase frequency. In fact this regime can be shown to be characterized by a single frequency $\omega_{\xi_1}^{\rm o}=\mu\hbar k$ even for a perfect overlap of all paths involved and is describable by an analogue form as (\ref{totint16}) and (\ref{totint17}) with $\vec{k}\vec{g}=0$ and $\mu\to\mu/2$.

As a final remark, we have assumed a symmetric momentum inversion in the intermediate $\tau$-region that does not lead to additional $\xi_1$-phase contributions. If that condition is not strictly fulfilled, all frequency contributions will be shifted by the corresponding phase, except for the closed paths contributions.


\section{Can quantum gravity effects be expected in interferometric setups of massive particles?}\label{sect_obs}
The question whether quantum gravitational effects can be expected in such a quantum optical interferometric setup turns out to be very controversial. Whereas an energy dispersion modification has been proposed in many theories to date\,\cite{alfaro00,magueijo02, amelino02}, its inclusion into the `test framework' of standard quantum mechanics\,\cite{mercati10} remains debatable. In light of the incompleteness and controversies of existing quantum gravity theories, probing such a small detail in a large framework seems a promising starting point, that can both help to validate and promote a better understanding of Planck scale physics. However one should keep in mind, that a theory of quantum gravity is much more than merely a modified energy dispersion relation or a modified commutator. It should in addition specify the underlying metric, the behaviour under transformations that might go along with a deformed Poincar\'e symmetry, a kinematic description of the equations of motion, and an interpretation of the physically relevant coordinates and observables\,\cite{hossenfelder13}.\\ 
The question remains whether massive particles are appropriate for testing quantum gravity. As for macroscopic bodies the proposed energy dispersion corrections would be large, what is unobserved on a macroscopic scale, it is a natural assumption to introduce a restriction to particle sizes that do behave `quantum mechanically'\,\cite{mercati10}. Note that the proposed energy dispersion corrections (\ref{pqg1}) constitute a perturbative expansion for small masses and would consequently not be valid in the macroscopic limit anyway. Even more delicate, a consistent framework would most probably require a modification of special relativity, and for a curved metric this will result in a non-linear momentum addition law. This can be motivated by the fact, that a minimal lengthscale maintains a fundamental role only if it is observer independent, contradictory to the standard framework of special relativity\,\cite{magueijo02, amelino02}. As a modified velocity addition law has been the consequence of defining a fundamental velocity in special relativity (the speed of light), it is not surprising that introducing in addition a fundamental lengthscale can lead to a modified momentum addition. Such a non-linear addition will however lead to a scaling problem as an iteration of a correction quadratic in the momentum scales quadratically with the number of constituents and will eventually become significantly large in contradiction to the `macroscopic world' observation. This is well-known in the literature as the `soccer ball problem'\,\cite{amelino11, magueijo06, hossenfelder13}, such that a restriction of the theory to `fundamental particles' has been stated by many authors, even though this notion remains imprecise. A possible solution for composite particles of $N$ constituents has been proposed by replacing the Planck mass $M_p$ by $N\,M_p$\,\cite{amelino11, magueijo06}; an approach that would not result in any advantages for nanoparticles over atoms in testing quantum gravity.
Last, there have been some proposals, that the quantum gravity corrections depend on the mass density rather than the absolute particle mass\,\cite{hossenfelder07}. 
Remarkably, apart from all those controversies, a test of the energy dispersion relation, will serve a test for the validity of special relativity even in a broader context\,\cite{mattingly05, amelino09}.

\begin{figure}[htb]
\begin{centering}
\includegraphics[scale=0.4]{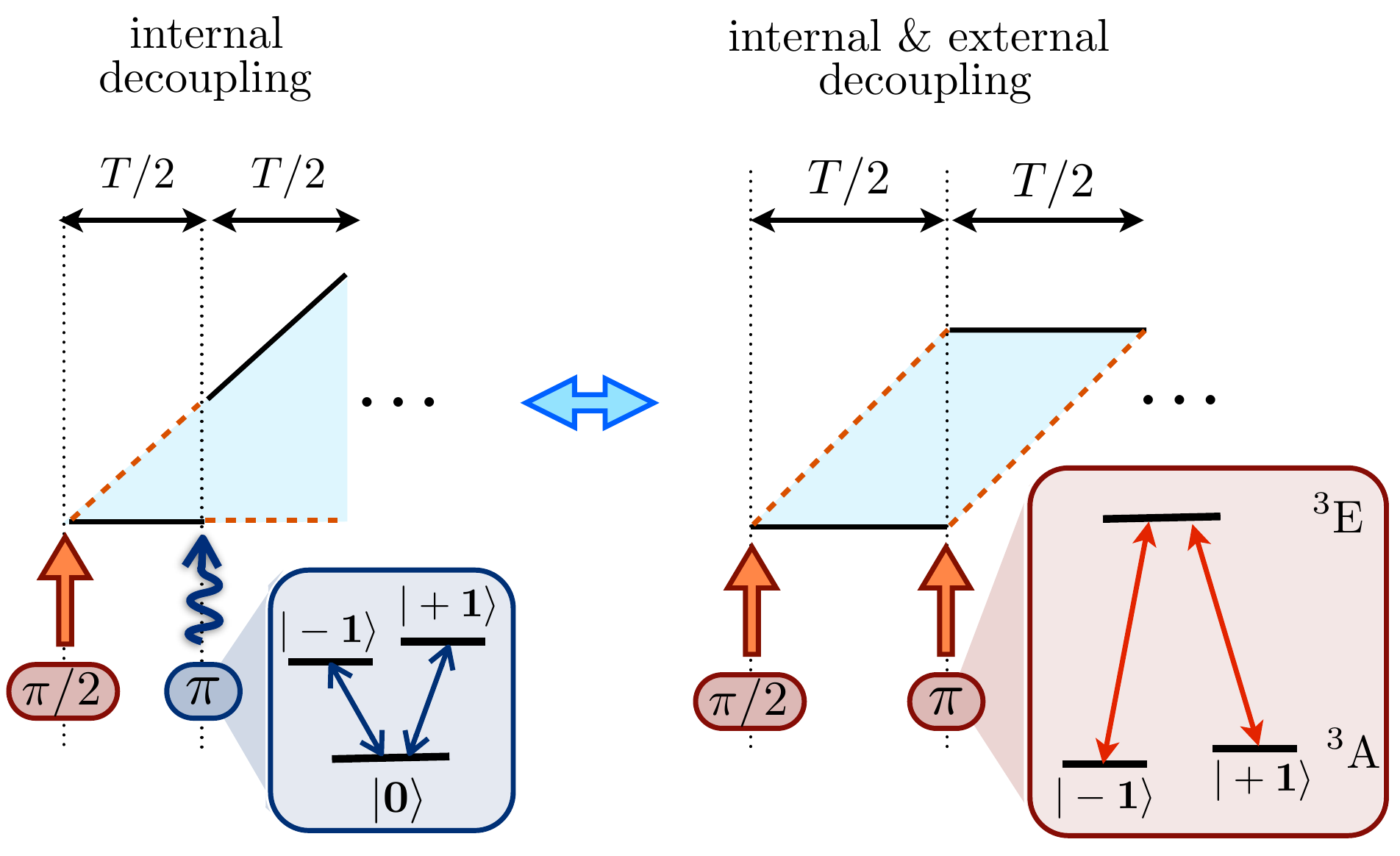}
\caption{\label{b_decoupl} (Color online) \textbf{Pulsed decoupling scheme. } \textit{Left:} Internal decoupling by a recoil free (microwave-) $\pi$-pulse. \textit{Right:} Decoupling of the internal and external degrees of freedom by a Raman-(recoil)-laser pulse, leading to an accelerometer (gravimeter) interferometry setup. Red arrows indicate Raman laser transitions whereas blue wavy arrows represent microwave transitions. Black and red dashed lines represent the internal ground and excited state, respectively. 
 }
\end{centering}
\end{figure}

\begin{figure*}[!ht]
\begin{centering}
\includegraphics[scale=0.45]{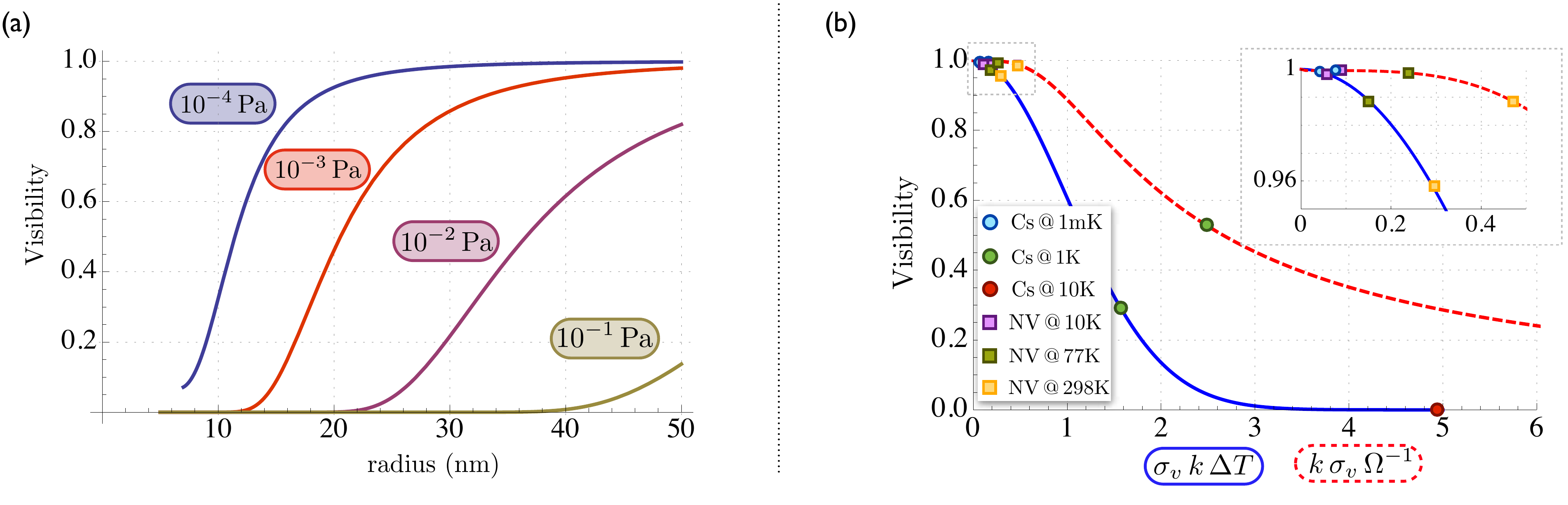}
\caption{\label{b_deceffects} (Color online) \textbf{(a) Collisional decoherence:} Interferometer visibility for a (background gas) temperature of 10\,K and different nanodiamond radii and pressures as indicated in the figure. The total interferometer time has been chosen as $2\,T=200\,\mu s$ and the interferometric path splitting is assumed constant and approximated by its maximal value $\Delta z=\hbar k^2/m\,T$ with $k=2\cdot 10^7\,{\rm m}^{-1}$. An isotropic elastic scattering cross section has been assumed in a description as developed in\,\cite{hornberger03}.  \textbf{(b) Time and pulse errors:}  Visibility reduction for time (\textit{blue solid}) and pulse (Doppler shift) errors (\textit{red dashed}) for different thermal velocity variances $\sigma_v$. Values for Cs atom and nanodiamond (radius 5\,nm) setups at different temperatures are indicated in the figure, assuming $\Omega=2\pi\cdot 10{\,\rm MHz}$ and  $\Delta T=10\,{\rm ns}$, respectively.   }
\end{centering}
\end{figure*}

\section{Interferometry with massive particles and decoherence}\label{sect_pract}
Here we briefly discuss practical issues of such an interferometric implementation based on nanoparticles: The decoupling from decoherence, collisional decoherence, imperfect pulses and time constraints and their impact on the interference fringe visibility. We will focus here on the robust stability regime; however as these effects are either general or mainly based on the zeroth order phase term, these results also hold true for other interferometric applications with nanoparticles and even the same scaling properties can be expected for different types of interferometers.

Decoherence decoupling schemes\,\cite{delange10, *naydenov11, *cai12} for the internal degrees of freedom, crucial to reach significant $T_2$-coherence times and in addition removing quasi-static energy shifts, can be implemented by (almost) recoil free microwave $\pi$-pulses within the ground state triplet, as shown in figure\,\ref{b_decoupl}. Note that recoil based $\pi$-pulses, obtained in analogy to the beam splitter interactions, would lead to the well-known gravimeter configuration\,\cite{kasevich91, kasevich92, schleich2013}. This latter setup, unsuitable for the anticipated task, reveals a decoupling in both the external and internal degrees of freedom, which makes it only sensitive to accelerations ($V'(x)\neq 0$). 

Figure\,\ref{b_deceffects} analyzes several sources of decoherence and visibility loss. Following the fact that for recoil based beam splitters the space-time area scales inversely with the particle mass $\propto 1/m$ and that the thermal velocity variance $\sigma_v\propto 1/\sqrt{m}$, these decoherence effects show a favorable scaling with increasing mass. Background gas collision induced decoherence, the prominent spatial decoherence source\,\cite{hornberger12} for nanoparticles, is illustrated in figure\,\ref{b_deceffects}\,(a)  for different pressures and particle sizes. Here a gas particle momentum change $\Delta\vec{p}$ during the scattering process will lead to a (random) phase $\Delta\phi=(1/\hbar)\Delta\vec{p}\,\Delta\vec{x}$ with $\Delta\vec{x}\propto 1/m$ the interferometer path separation. The effects of time imprecisions and Doppler shift induced imperfect population transfer are depicted in figure\, \ref{b_deceffects}\,(b). A deviation of the interferometer cycle times by $\Delta T$ will lead to an additional momentum dependent phase term subject to thermal decoherence $\phi_{o}=(\vec{p}/m)\,\vec{k}\,\Delta T=\vec{v}\,\vec{k}\,\Delta T=\vec{p}/\hbar\,\Delta \vec{x}$ with $\vec{v}=\vec{p}/m$ the particle velocity. This defines a minimal required time precision that scales inversely with the particle mass. Similarly, a Doppler shift detuning $\delta\sim \vec{v}\vec{k}$ will affect the population transfer of the interferometric beam splitter operations, an effect that is not removed by introducing decoupling sequences. Owing to the velocity variance mass scaling, its influence also decreases with increasing particle size. In a more general context, best seen in perturbative path integral approaches for the interferometer phase calculation\,\cite{storey94}, smaller space-time areas  reduce the phase contributions of mass independent spatial Hamiltonian perturbations.

Note however that, despite the advantages outlined above, from a practical perspective the interference with nanoparticles is much more complex compared to atomic setups (see also section\,\ref{sect_pb}). In addition this decoherence scaling only holds for nanoparticles, as going to larger sizes would lead to significant other decoherence contributions like photon scattering and thermal decoherence\,\cite{hornberger12} and would eventually end up in classical behaviour.

\section{Conclusion}
In this paper we have shown how interferometry with massive particles, with the specific examples of nanodiamonds, can improve existing bounds on proposed quantum gravity corrections to the energy dispersion relation. Assuming the Planck scale as the relevant scale for quantum gravity, this would even allow for an existence proof of the linear correction contribution under experimentally realistic parameters. However, despite its rather large proposed magnitude under optimal conditions, gravitation and a thermal distribution of momenta will in general render its observation inaccessible, which necessitates a revision of previously estimated bounds. As the requirements on the particle temperature and gravitation in a standard setup turn out to be quite challenging, we have proposed an alternative noise insensitive scheme based on an initial momentum transfer combined with a momentum inversion by gravitation. Moreover, we have investigated the influence of the individual interferometer paths on the total interference pattern along with its dependence on  different temperatures and nanodiamond sizes. 
Last, the combination of the interferometric setup with decoupling sequences and mass scalings for decoherence and visibility reducing processes have been analyzed. The latter reveal an increasing robustness with the particle size, originating in decreasing spatial areas for increasing masses. This behaviour renders the interference of nanoparticles, despite technical challenges, a viable scenario.   

\begin{acknowledgments}
This work was supported by the Alexander von Humboldt Foundation, the EU STREP project EQuaM and the EU Integrating projects SIQS and DIADEMS.
\end{acknowledgments}


\bibliography{interfer_bib}


\makeatletter
\renewcommand*{\p@section}{}
\renewcommand*{\p@subsection}{}
\renewcommand*{\p@subsubsection}{}
\makeatother

\setcounter{figure}{0}
\setcounter{equation}{0}
\renewcommand{\thefigure}{A\arabic{figure}}
 \renewcommand{\thetable}{A\arabic{table}}
\renewcommand{\theequation}{A.\arabic{equation}}
\noindent

\appendix
\onecolumngrid
\section{Interferometer phase in the Ramsey Bord\'e setup}\label{append_1}
\begin{figure}[htb]
\begin{centering}
\includegraphics[scale=0.45]{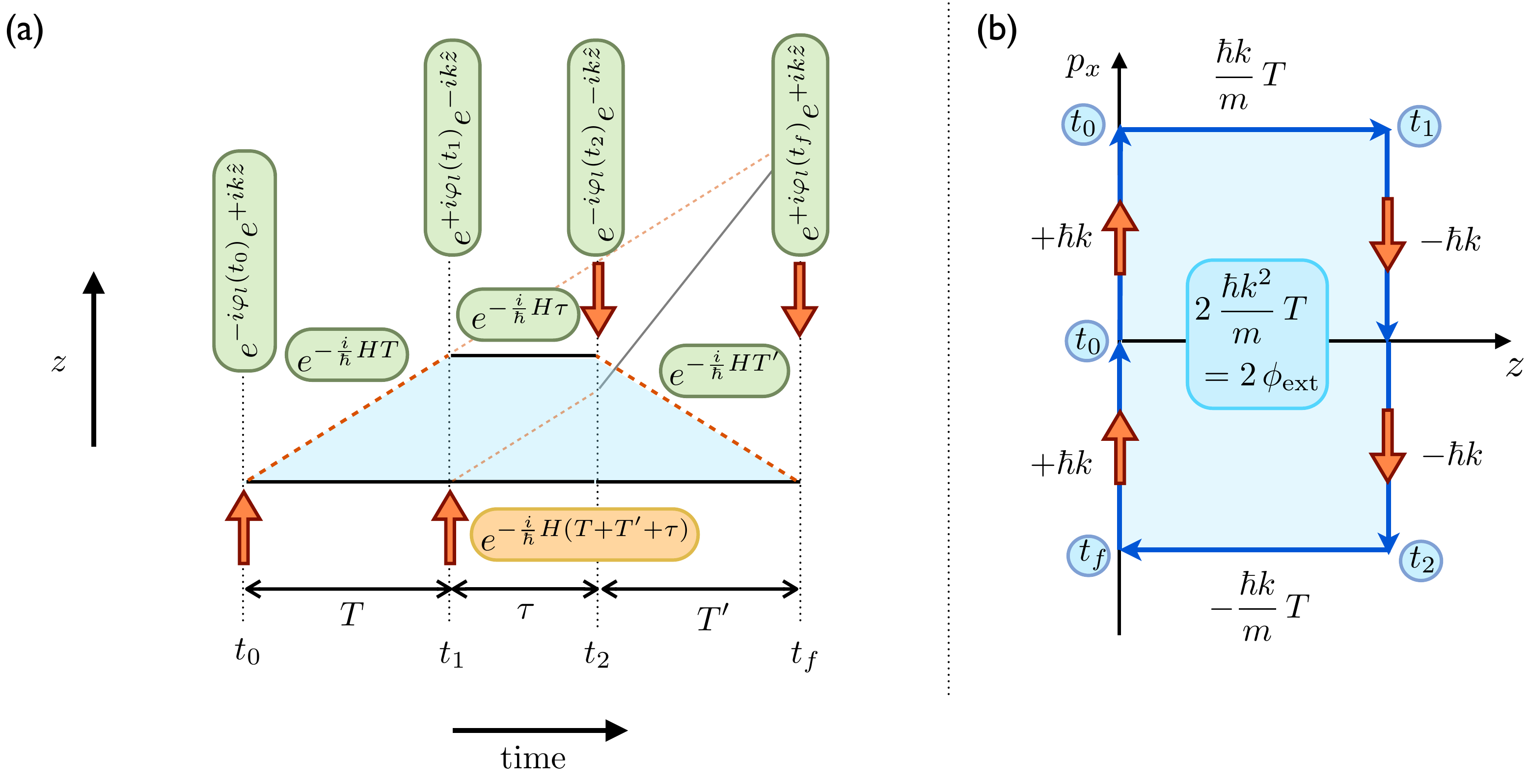}
\caption{\label{b_rbphase} (Color online) \textbf{Ramsey Bord\'e interferometer and phase space area} \textbf{(a)} Ramsey Bord\'e interferometry setup. Black and red dashed lines correspond to the internal states $\ket{g}$ and $\ket{e}$, respectively. Red arrows indicate $\pi/2$-laser pulses in the corresponding direction. The unitary evolution operators for the upper and lower path, leading to $U^{(g)}$ as defined in the main text, are shown in green and orange boxes, respectively. \textbf{(b)} Phase space evolution and area for $E(p)=p^2/(2m)$ in the absence of an external potential. Laser interaction paths are marked by the superimposed red arrow and in addition the corresponding times are given under the assumption of instantaneous laser pulses.}
\end{centering}
\end{figure}
In here we will derive the interferometric phase of the Ramsey Bord\'e setup for the lower closed paths configuration as depicted in figure\,\ref{b_rbphase}(a) using the operator based formalism developed in\,\cite{schleich2013}. Thereby we will focus on keeping the analysis as general as possible, what allows for a simple inclusion of a modified energy dispersion relation or accelerating inertial forces at a later stage. From this specific example we will obtain a general rule for the interferometric phase calculation based on the kinetic energy, in cases when the external potential is at most linear in position (inertial force), which  allows for the phase calculation of any arbitrary path combination as performed in section\,\ref{sect_totint}. 

We will assume the two level system $\{ \ket{g}, \ket{e} \}$ initially prepared in the state $\ket{\psi_i}=\ket{g}\otimes\ket{\psi_{\rm ext}}$, with $\ket{\psi_{\rm ext}}$ describing the external, motional degrees of freedom. The $\pi/2$- laser interaction pulse, that takes the role of a beam-splitter, is described by the unitary operation
\begin{equation}\label{ip1} U_{\pi/2} =\frac{1}{\sqrt{2}}\,\left( \mathbbm{1}-i\,\left[e^{-i\,\varphi}\,e^{i\,k\,\hat{z}}\sigma_++e^{i\,\varphi}\,e^{-i\,k\,\hat{z}}\sigma_-  \right] \right) \end{equation}
with the rotation axis defined by $\sigma_\varphi=\cos(\varphi)\,\sigma_x+\sin(\varphi)\,\sigma_y$. Note that, most importantly, the recoil contribution $\exp(ik\xo)$ leads to a splitting in momentum space, which is responsible for the relevant splitting in the external degrees of freedom (`interferometric beam splitter operation'). During the free evolution periods, the external degrees of freedom evolve according to
\begin{equation}\label{ip2} U_{f}=\exp\left( -(i/\hbar)\,H(\hat{p},\hat{z})\,t \right)\quad\text{with} \quad H(\hat{p},\hat{z})=E(\hat{p})+V(\hat{z})\,\quad \text{and} \quad V'(\hat{z})=\text{const}.  \end{equation} 
Herein $E(\hat{p})$ denotes the kinetic energy and $V(\hat{z})$ an external potential; importantly we restrict this to potentials linear in position, as only in that case a simple closed form for the phase can be obtained. Note however that for the case of quadratic harmonic oscillator potentials, a straightforward phase expression can be obtained as well by a slightly different approach\,\cite{vacanti12, scala13}. The internal degrees of freedom evolution follows out of
\begin{equation}U_{\rm in}=\exp\left(-(i/\hbar)\,\int^t\,H_{\rm in}(t')\,\mathrm{d}t'  \right) \quad \text{with} \quad H_{\rm in}=-\hbar\,\delta(t)/2\,\sigma_z \end{equation}
in a frame rotating with the beam splitter laser frequency, $\delta(t)$ the corresponding detuning and $\sigma_z$ the Pauli z-matrix defined in the two level internal state system. This term can be included conveniently in (\ref{ip1}) by replacing $\varphi\rightarrow \varphi_l(t)=\varphi(t)+\int^t\delta(t')\mathrm{d}t'$.
Right after the interferometer sequence, and restricting to the two paths involved in the sequence, i.e. the ones that lead to equal states, the system ends up in the state
\begin{equation}\label{ip3}  \ket{\psi_f}=\left( \frac{1}{\sqrt{2}} \right)^4\,\left( \left[ U_u^{(g)}+U_l^{(g)} \right]\,\ket{\psi_i}+i\,\left[ U_u^{(e)}-\,U_l^{(e)} \right]\,\ket{\psi_{i}} \right)  \end{equation}
with $U_u$ and $U_l$ describing the evolution operators of the upper and lower path, respectively, and the left part evolution ends up in the atomic state $\ket{g}$, whereas the right one ends up in $\ket{e}$ as indicated by the indices $g$ and $e$, respectively.  \\
The `closed path' contribution to the probability for finding the system in the internal state $\ket{g}$ (=the initial state) after the interferometer sequence follows out of\,(\ref{ip3}) and is given by 
\begin{equation}\label{ip5} p_g'=\text{tr}\left( \left[ \ket{g}\bra{g}\otimes \mathbbm{1} \right]\,\ket{\psi_f}\bra{\psi_f} \right)=\frac{1}{8}\,\left[ 1+\frac{1}{2}\,\left(\bra{\psi_i}U_l^{(g)\dagger}\,U_u^{(g)}\ket{\psi_i}+\text{c.c.} \right) \right]\,.\end{equation}
For an initial mixed state $\rho_{\rm in}$ the interference term $\bra{\psi_i}U_l^{(g)\dagger}\,U_u^{(g)}\ket{\psi_i}$ has to be replaced by $\text{tr}(U_l^{(g)\dagger}\,U_u^{(g)}\rho_{\rm in})$. 

Thus it remains to calculate $U_l^{(g)\dagger} U_u^{(g)}$, a quantity that delivers a pure phase for the closed interferometer path combinations here, making it independent of the specific form of the initial state $\ket{\psi_i}$. Note that, as both paths do end up in the same state, both evolutions are formally equivalent up to an operator ordering, i.e. the non-commutativity of operators is responsible for the appearance of an interferometric phase. \vspace{2ex}

\subsubsection*{Important operator relations for the evaluation of the unitary path evolution operators}
For the calculation of the unitary path evolution operators it is important to recall some operator relations, obtained out of the operator identity $\exp(-\alpha \hat{A})\hat{B}\exp(\alpha \hat{A})=\sum_{\nu=0}^\infty (-\alpha)^\nu/\nu!\,[\hat{A},\hat{B}]_\nu$ with $[\hat{A},\hat{B}]_0=\hat{B}$ and $[\hat{A},\hat{B}]_\nu=[\hat{A},[\hat{A},\hat{B}]_{\nu-1}]$ along with $[\hat{z},\hat{p}]=i\hbar$. Moreover as mentioned before, we will assume the potential being at most linear in position, such that $V'(\hat{z})=\text{const}$.\\
Then it follows that
\begin{equation}\label{ip6}  \e^{\pm ik\,\hat{z}}\,\e^{-i\,\phi(\hat{p})}\,\e^{\mp i k \hat{z}}=\e^{-i\phi(\hat{p}\mp \hbar k)} \quad \text{and} \quad \e^{\frac{i}{\hbar} V(\hat{z})\,t}\,\e^{-i\phi(\hat{p})}\,\e^{-\frac{i}{\hbar} V(\hat{z})\,t} =\e^{-i\,\phi(\hat{p}-V'(\hat{z})\,t)} \end{equation}
where we have used in addition that $[V(\hat{z}),\hat{p}]=V'(\hat{z})\,[\hat{z},\hat{p}]$. This represents the momentum gain by the beam splitter operation and the external potential, respectively. Moreover the evolution sequence calculation is significantly simplified by separating the position and momentum operator contributions in the evolution (\ref{ip2}), namely 
\begin{equation}\label{ip30}  \e^{-\frac{i}{\hbar}\,H(\hat{p},\hat{z})\,t}=\e^{-\frac{i}{\hbar}\,V(\hat{z})\,t}\,\e^{-\frac{i}{\hbar}\int_0^t E(\hat{p}-V'\,t')\,\mathrm{d}t'}\,,  \end{equation}
where $V'\equiv V'(\hat{z})$ and the second contribution corresponds to the interaction picture evolution with respect to the potential, i.e. $\exp(-(i/\hbar) \int H_{\rm int}(t') \mathrm{d}t')$ with $H_{\rm int}(t)=\exp[(i/\hbar)V(\hat{z}) t]\,E(\hat{p})\,\exp[-(i/\hbar) V(\hat{z}) t]$, whereas the first contribution accounts for the back-transformation to the original frame.

\subsubsection*{Path evolution operators and interference phase calculation}
We will now turn to the calculation of the evolution operators, that are given by (see figure\,\ref{b_rbphase})
\begin{equation}\begin{split}\label{ip10} U_u^{(g)}&=\e^{i\,\Delta\varphi} \,\,\e^{i\,k\,\xo}\,\e^{-i/\hbar\,H\,(t_f-t_2)}\,\e^{-i\,k\,\xo}\,\e^{-i/\hbar\,H\,(t_2-t_1)}\,\e^{-i\,k\,\xo}\,\e^{-i/\hbar\,H\,(t_1-t_0)}\,\e^{i\,k\,\xo} \\
U_l^{(g)}&=\e^{-i/\hbar\,H\,(t_f-t_0)}
 \end{split} \end{equation}
with the laser and internal phase contributions
\begin{equation}\label{ip10}  \Delta\varphi=\left[ \varphi_l(t_f)-\varphi_l(t_2)+\varphi_l(t_1)-\varphi_l(t_0) \right]=\left[ \varphi(t_f)-\varphi(t_2)+\varphi(t_1)-\varphi(t_0) \right]+\int_{t_0}^{t_1}\delta(t')\,\mathrm{d}t'+\int_{t_2}^{t_f}\delta(t')\,\mathrm{d}t' \end{equation}
that does take the value $\Delta\varphi=2\,\delta\,T$ for fixed laser phases and a constant detuning $\delta(t)\equiv \delta$. \\
Substituting the free evolution sequences by (\ref{ip30}) and commuting the position operator dependent terms to the left with the help of (\ref{ip6}) the evolution sequence can be simplified to (again assuming that $V'(\hat{z})$ is independent of $\hat{z}$)
\begin{equation} \label{ip15} \begin{split}
U_{u,{\rm int}}^{(g)}&=\e^{i\,\Delta\varphi}\,\e^{-\frac{i}{\hbar}\int_{t_2}^{t_f}E(\hat{p}-\hbar k-V'\,t')\mathrm{d}t'}\,\e^{-\frac{i}{\hbar}\int_{t_1}^{t_2}E(\hat{p}-V'\,t')\mathrm{d}t'}\,\e^{-\frac{i}{\hbar}\int_{t_0}^{t_1}E(\hat{p}+\hbar k-V'\,t')\mathrm{d}t'}=\e^{i\Delta\varphi}\,\e^{-\frac{i}{\hbar}\int_{t_0}^{t_f} E(\hat{p}_u(t'))\mathrm{d}t'}\\
U_{l,{\rm int}}^{(g)}&= \e^{-\frac{i}{\hbar}\int_{t_0}^{t_f}E(\hat{p}-V'\,t')\mathrm{d}t'}=\e^{-\frac{i}{\hbar}\int_{t_0}^{t_f} E(\hat{p}_l(t'))\mathrm{d}t'} ,
\end{split}  \end{equation}
where in the last step we used that the external phase part corresponds just to the kinetic energy along the path. Here $\hat{p}(t')$ is the classical momentum in time along that path with the initial momentum replaced by the operator $\hat{p}$. Therefore the operator product appearing in the interference term\,(\ref{ip5}) is given up to the laser phase contributions by the kinetic energy integrated along the closed path
\begin{equation}  U_l^{(g)\dagger}\,U_u^{(g)}=\e^{i\,\Delta\varphi}\,\exp\left(-\frac{i}{\hbar}\oint E[\hat{p}(t')]\,\mathrm{d}t'\right)\,. \end{equation} 

This phase calculation, following the same steps as above, can be generalized to arbitrary paths involved, resulting in
\begin{equation} \label{ip32} U_j=e^{i\varphi_{\rm laser}}\,e^{\frac{i}{\hbar} p_{\rm f}^j\,\hat{z}}\,e^{-\frac{i}{\hbar}\,\int_{t_0}^{t_f} E(\hat{p}_j(t'))\,\mathrm{d}t'}  \end{equation}
for a path $j$ with $p_{\rm f}^j$ the effective total momentum gain, $\hat{p}_j(t)$ the momentum along that path and $\varphi_{\rm laser}$ the laser phase internal contribution. More generally, this holds true for any interferometer subject to laser pulses and under a potential at most linear in position and has been used previously for the phase evaluation based on a semiclassical treatment e.g. in\,\cite{cadoret09}.

As an explicit example let us consider the closed paths combination as above subject to the standard energy dispersion relation $E(\hat{p})=\hat{p}^2/(2m)$, a gravitational field $V(\hat{z})=m\,\vec{g}\,\vec{x}$ and assuming that the laser phase is kept constant as well as the detuning. Noting that in the above derivation only the force component parallel to the pulse direction does contribute (the non-commuting $\hat{p}$ and $\hat{z}$ components), one obtains the well-known expression for the phase ($T=T'$)
\begin{equation}\label{ip17} \phi=\frac{\hbar\,k^2}{m}\,T+\vec{k}\,T\,\left[\vec{g}\,(T+\tau)-\frac{1}{m}\Delta\vec{p}_{\rm acc}\right]+2\,\delta\,T  \end{equation}
where we introduced an additional acceleration of momentum $\Delta\vec{p}_{\rm acc}$ during the evolution time $\tau$, besides the gravitational influence, often used in experiments to increase the signal by adding a series of N acceleration pulses such that $\Delta\vec{p}_{\rm acc}=-N\,\hbar \vec{k}$.

\subsection{Relation to the phase space area}
As previously noted by many authors, the motional part of the interferometer phase $\phi_{\rm ext}$ is, in many cases, related to the phase space area $a_{\rm ps}$ by the relation
\begin{equation}\label{ps_rel} a_{\rm ps}=\oint p\,\mathrm{d}z=2\,\hbar \phi_{\rm ext}\,, \end{equation}
which seems to hold for $E(p)=p^2/(2m)$ by noting that in that case the phase space area (in the absence of any acceleration) is given by $a_{\rm ps}=2\,(\hbar k^2)/m\,T$ (see figure\,\ref{b_rbphase}(b)). Moreover such a relation can be strictly proven for trapped particle interferometers, in which the interferometer action can be described by a series of continuous displacements in phase space\,\cite{vacanti12}. For the case considered here, as has been shown in\,\cite{schleich2013}, the phase space area can be related to the kinetic energy by noting that $\mathrm{d}z=(i/\hbar) [H,z]\mathrm{d}t=(i/\hbar) [E(p),z]\mathrm{d}t=(\mathrm{d}/\mathrm{d}p) E(p)\,\mathrm{d}t$
\begin{equation}\label{ip19} a_{\rm ps}=\oint p\,\mathrm{d}z=\oint p\,\frac{\mathrm{d}}{\mathrm{d}p}E(p)\,\mathrm{d}t\overset{?}{=}2\,\hbar\,\phi_{\rm ext}=2\oint\mathrm{d}t\, E(p)\,.  \end{equation}
Thus, the relation of the phase space area to the external interferometric phase as given in (\ref{ps_rel}) follows naturally if two conditions are fulfilled: (i) the energy dispersion relation is of the standard form $E(p)=p^2/(2m)$ and (ii) the system is at most subject to inertial forces for the phase energy relation to be valid.

\setcounter{figure}{0}
\setcounter{equation}{0}
\renewcommand{\thefigure}{B\arabic{figure}}
 \renewcommand{\thetable}{B\arabic{table}}
\renewcommand{\theequation}{B.\arabic{equation}}

\section{Momentum overlap matrix element for a thermal harmonic oscillator state}\label{append_overlap}
In this appendix we will derive the coherence matrix element $\ex{p|\rho_{\rm th}|p+\Delta p}$, that accounts for the finite state overlap in the interference term calculations (\ref{totint6}). This matrix element defines the visibility decay for paths open in position or momentum space. We will assume the particle starting initially in a thermal harmonic oscillator state,  that is  $\rho_{\rm th}=1/\mathcal{Z}\,\exp(-\hat{H}/(k_B\,\mathcal{T}))$ with $\hat{H}=\hbar\omega (\hat{n}+1/2)$, $\hat{n}=\hat{a}^\dagger \hat{a}$ the number operator and the partition function $\mathcal{Z}=\text{tr}[\exp(-\hat{H}/(k_B\,\mathcal{T}))]$.

Using that $\exp\left((i/\hbar)\Delta p \,\hat{z}\,\right)\ket{p}=\ket{p+\Delta p} $ along with the Fourier representation of the $\delta$-function $\delta(p-\hat{p})=1/(2\pi\hbar)\,\int_{-\infty}^\infty\exp(i/\hbar (p-\hat{p})y)\,\mathrm{d}y$, the matrix element can be rewritten as
\begin{equation}\label{tm5}  \ex{p|\rho_{\rm th}|p+\Delta p}=\ex{e^{\frac{i}{\hbar}\Delta p\,\hat{z}}\,\delta(p-\hat{p})}_{\rho_{\rm th}}=\frac{1}{2\pi\hbar}\,\int_{-\infty}^\infty\mathrm{d}y\,e^{\frac{i}{\hbar}p\,y}\,\ex{e^{\frac{i}{\hbar}\Delta p\hat{z}}e^{-\frac{i}{\hbar}y\hat{p}}}_{\rho_{\rm th}} \,. \end{equation}
We will start by evaluating the thermal expectation value, that upon application of the Baker-Campbell Hausdorff theorem can be cast into the form
\begin{equation}\label{tm6}  \ex{e^{\frac{i}{\hbar}\Delta p\,\hat{z}}e^{-\frac{i}{\hbar}y\,\hat{p}}}_{\rho_{\rm th}} =e^{\frac{i}{\hbar}(\Delta p/2)  y}\,\left\langle e^{\frac{i}{\hbar}[\Delta p\,\hat{z}-y\,\hat{p}]} \right\rangle_{\rho_{\rm th}}=e^{\frac{i}{\hbar}(\Delta p/2)  y}\,\left\langle e^{-\frac{i}{\hbar}\left[ (\gamma+i\delta)\,a^\dagger+(\gamma-i\delta)\,a \right]}\right\rangle_{\rho_{\rm th}} \end{equation}
where in the last step we have used that $\hat{z}=\sqrt{\hbar/(2m\omega)}\,(a+a^\dagger)$ and $\hat{p}=i\sqrt{(\hbar m \omega)/2}\,(a^\dagger-a)$. Furthermore, for simplicity of the expressions, we have defined $\gamma\equiv -\sqrt{\hbar/(2m\omega)}\,\Delta p$ and $\delta\equiv \sqrt{(\hbar m \omega)/2}\, y$. Now, making use of
\begin{equation}\left\langle e^{\lambda \hat{a}^\dagger+\mu\,\hat{a}} \right\rangle_{\rho_{\rm th}}=e^{\frac{1}{2}\ex{(\lambda\hat{a}^\dagger+\mu\hat{a})^2}_{\rho_{\rm th}}} \end{equation}
 and noting that $\rho_{\rm th}$ is diagonal in the energy eigenbasis and thus only excitation number conserving terms have to be considered, expression\,(\ref{tm6}) can be evaluated to
\begin{equation}\label{tm7}  \ex{e^{\frac{i}{\hbar}\Delta p\hat{z}}e^{-\frac{i}{\hbar}y\hat{p}}}_{\rho_{\rm th}}=e^{\frac{i}{\hbar}(\Delta p/2)  y}\,e^{-\frac{1}{2\hbar^2}\,(\gamma^2+\delta^2)\,(2\ex{\hat{n}}+1)}=e^{\frac{i}{\hbar}(\Delta p/2)  y}\,e^{-\frac{1}{2\hbar^2} \ex{\hat{p}^2}\,y^2}\,e^{-\frac{1}{2\hbar^2} \ex{\hat{z}^2}\,\Delta p^2}  \end{equation}
with the momentum and position variance given by
\begin{equation}\label{tm8} \ex{\hat{p}^2}\equiv\sigma_p^2=\frac{m\omega\hbar}{2}\,\left( 2\ex{\hat{n}}+1 \right)\, ,  \qquad
						\ex{\hat{z}^2}\equiv\sigma_z^2=\frac{\hbar}{2m\omega}\,\left( 2\ex{\hat{n}}+1 \right)	\, .			
\end{equation}
and the thermal population $\ex{\hat{n}}=(\exp(\hbar\omega/(k_B\,\mathcal{T}))-1)^{-1}$.
Now inserting (\ref{tm7}) into (\ref{tm5}) and performing the $y$-integration, one obtains the final result
\begin{equation}\label{tm9} \ex{p|\rho_{\rm th}|p+\Delta p}=\frac{1}{\sqrt{2\pi\ex{\hat{p}^2}}}   e^{-\frac{1}{2\ex{\hat{p}^2}}\,\left(p+\frac{\Delta p}{2}  \right)^2}\,e^{-\frac{1}{2\hbar^2}\,\ex{\hat{z}^2}\,\Delta p^2}\,. \end{equation}
The first contribution corresponds to a Gaussian distribution in momentum, that will converge to the Boltzmann distribution in the high temperature limit $k_B\,\mathcal{T} \gg \hbar \omega$ where $\ex{p^2}\simeq m k_B \mathcal{T}$ and will turn out to characterize the decay for a final path separation of the interferometer arms in position. In addition, the finite overlap in momentum space is accounted for by the last term, that does represent a direct decay in $\Delta p$ on a characteristic momentum scale $[\hbar/\sqrt{\hat{z}^2}]$.

\setcounter{figure}{0}
\setcounter{equation}{0}
\renewcommand{\thefigure}{C\arabic{figure}}
 \renewcommand{\thetable}{C\arabic{table}}
\renewcommand{\theequation}{C.\arabic{equation}}

\section{The stability regime}\label{append_2}
\begin{figure}[htb]
\begin{centering}
\includegraphics[scale=0.45]{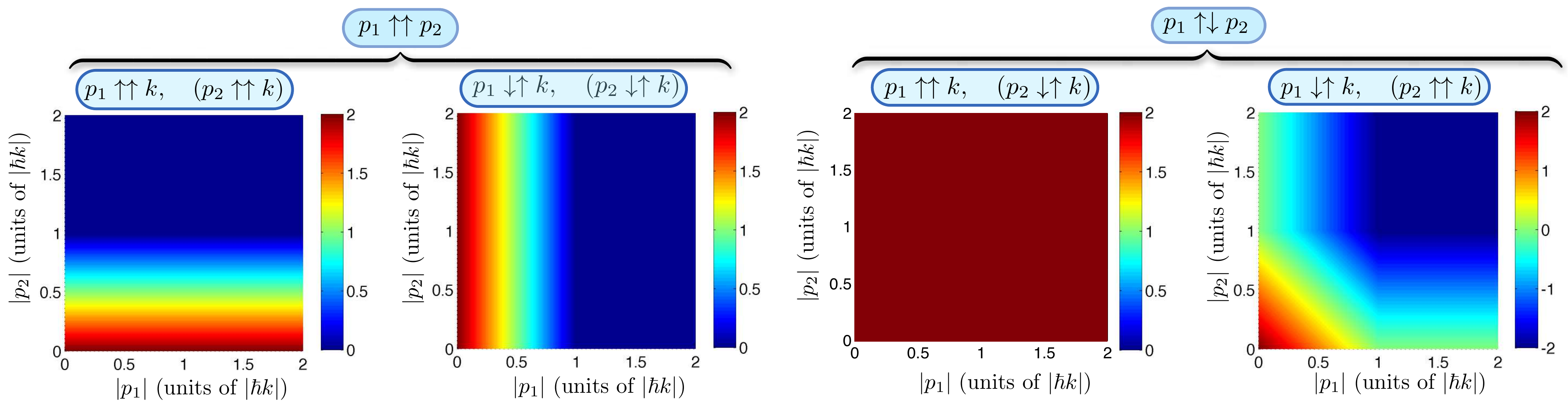}
\caption{\label{b_sup_stability} (Color online) \textbf{Phase contribution in the parallel regime.} Phase contribution $\phi/(2\mu\,|\hbar k| T)$ vs different momentum values $p_1$ and $p_2$. The third configuration corresponds to the stability regime. }
\end{centering}
\end{figure}
In the absence of specifically designed momenta directions, the $\xi_1$-phase contribution is most likely to decay in the presence of momenta exceeding the beam splitter recoil contribution, $|p|\geq |\hbar k|$. As shown in section\,\ref{sect_oo}, this leads to a suppression of coherent $\xi_1$-phase oscillations  under the influence of gravitation and thermal momentum distributions. However based on an analysis in the regime $|p|\gg|\hbar k|$\,(\ref{xi3}) it turned out that a change in the momentum with respect to the recoil direction in between the two interferometric cycles leads to a significant unsuppressed phase term, and more precise in a stability configuration $ \vec{p}_1^{\,0}\uparrow\uparrow \vec{k} \quad \&\&\quad \vec{p}_2^{\,0}\uparrow\downarrow \vec{k}\,\,(\Leftrightarrow  \vec{p}_1^{\,0} \uparrow\downarrow \vec{p}_2^{\,0})  $ the optimal phase $\phi=2\mu\hbar k T$ is recovered independently of the momentum value. Here, as already outlined in the main text, `parallel' and `orthogonal' refers to the direction of the momentum relative to the first recoil $\vec{k}$ direction; the first interferometric cycle is defined by $[t_0,t_1]$ and the second one by $[t_2,t_f]$ (see figure\,\ref{b_setup1}).  We will demonstrate here, that this does not only hold in the regime of large momenta, but in any possible limit as long as the parallel condition is fulfilled.\par
An orthogonal component will always lead to a decay with increasing momentum
and that even holds for the individual cycles, i.e. $p_1$ and $p_2$ components alone. In contrast, for a parallel component only the combination of both will lead to a potential suppression. Figure\,\ref{b_sup_stability} analyzes this effect in more detail for static, but potentially different parallel momenta $p_1=|\vec{p}_1^{\,0}|$, $p_2=|\vec{p}_2^{\,0}|$ for the two interferometer cycles. It turns out that, without a momentum direction change ($\vec{p}_1\uparrow\uparrow \vec{p}_2$), the phase contribution always decays to zero in the large momentum limit $|p_1|, |p_2|\gg |\hbar k|$, and therefore the only solution consists of working in the rather challenging $|p_1|, |p_2|\ll |\hbar k|$ regime. In contrast, in a $\vec{p}_1\uparrow\downarrow \vec{p}_2$ configuration, i.e. if there occurs a change in the momentum direction, the large momentum limit is characterized by a non-zero, momentum independent phase. Particularly, the stability regime $ \vec{p}_1^{\,0}\uparrow\uparrow \vec{k} \quad \&\&\quad \vec{p}_2^{\,0}\uparrow\downarrow \vec{k}\,\,(\Leftrightarrow  \vec{p}_1^{\,0} \uparrow\downarrow \vec{p}_2^{\,0}) $ retains this property in any possible momentum regime, as long as the parallel condition is fulfilled.

\setcounter{figure}{0}
\setcounter{equation}{0}
\renewcommand{\thefigure}{D\arabic{figure}}
 \renewcommand{\thetable}{D\arabic{table}}
\renewcommand{\theequation}{D.\arabic{equation}}

\section{Modified dispersion relation vs modified commutation relations}\label{append_3}
It is widely assumed that a quantization of space-time in quantum gravity will lead to a minimal length-scale of the order of the Planck length $L_p=\hbar/(M_p\,c)$, that can be accounted for by a modification of the commutation relation of the form\,\cite{hossenfelder13}
\begin{equation} [\hat{x},\hat{p}]=C(\hat{p})  \end{equation}
with $C(\hat{p})=i\hbar$ in the standard case of quantum mechanics and e.g. $C(\hat{p})=i\hbar\left(1+\xi\,p^2/(M_p\,c)^2 \right)$, that will lead to a minimal lengthscale $\Delta x=\sqrt{\xi}\,L_p$ according to the general uncertainty principle $\Delta x\Delta p\geq-1/2\,i\ex{C(\hat{p})}$\,\cite{hossenfelder13, pikovski12, ali11}. Herein $\xi$ denotes a dimensionless parameter that for a Planck-scale correction will be of order one and is in general upper bounded by the electroweak length scale to $\xi\leq10^{34}$\,\cite{ali11, das08}. Such a modified commutation relation will then lead to a modified interferometer phase, that can be calculated, at least in a perturbative way with respect to the Planck scale correction, in the formalism of Appendix\,\ref{append_1}, and its measurement has already been proposed in a more simplistic setup in\,\cite{pikovski12}. As the interferometer phase considered here corresponds merely to the kinetic energy integrated along a closed path, it is a natural question to ask if both, the modified energy dispersion relation and the modified commutation relation, are equivalent. That is, can the modified energy dispersion be reproduced by choosing the standard dispersion $E(\hat{p})=\hat{p}^2/(2m)$ and modifying the commutator instead (in particular with respect to (\ref{ip6}) and its appearance in the phase\,(\ref{ip15}))?
For this purpose we will begin by defining the energy dispersion relation via the displacement operation as
\begin{equation}\label{ue3} E(\tilde{p})=\left\langle p=0\left|\e^{-\frac{i}{\hbar} \tilde{p} \hat{x}}\,\frac{\hat{p}^2}{2m}\, \e^{\frac{i}{\hbar} \tilde{p} \hat{x}}\right|p=0 \right\rangle \end{equation} 
that can be evaluated (see also Appendix A and using that $[x,f(\hat{p})]=f'(\hat{p})[x,p]$) to
\begin{equation}\label{ue4} E(\tilde{p})=\left\langle p=0\left|\frac{1}{2m}\,\left[ \hat{p}^2-2\frac{i}{\hbar}\tilde{p}\,\hat{p}C(\hat{p})-\frac{\tilde{p}^2}{\hbar^2}\,C(p)^2+i\,\frac{\tilde{p}^3}{\hbar^3}\,C'(\hat{p})\,C(\hat{p})^2+\mathcal{O}(\tilde{p}^4) \right] \right|p=0 \right\rangle \end{equation}
which corresponds to an expansion in the Planck scale correction. For the modification of the commutation relation given above, this would lead to (with $\beta=\xi/(M_p c)^2$)
\begin{equation}\label{ue5} E(\tilde{p})=\frac{1}{2m}\,\left(  \tilde{p}^2+\frac{2}{3}\,\beta\,\tilde{p}^4+\frac{17}{45}\,\beta^2\tilde{p}^6+\mathcal{O}(\beta^3)\right)\,.  \end{equation}
It is worth noting that, due to the quadratic form of the unmodified energy dispersion relation, a correction term linear in the momentum does not appear according to\,(\ref{ue3}), whereas the absence of higher order odd powers in (\ref{ue5}) is only due to the specific form chosen for the commutator and in contrast to the absence of a linear term non-fundamental. Thus a quadratic correction could be reproduced by a modified commutator whereas a linear correction term is impossible within that framework. As a remark, such a conclusion is only valid if $\hat{p}$ corresponds to the physical momentum; if in contrast $\hat{p}$ is a function of the physical momentum operator as often used in quantum gravity approaches\,\cite{hossenfelder13}, a linear term correction would not be forbidden any more however related to a different coordinate framework of quantum mechanics.

\end{document}